\documentclass[12pt]{article}
\pdfoutput=1
\usepackage{putex}
\usepackage{amssymb}
\usepackage{amsmath}
\usepackage{empheq}
\usepackage{epsf}%\usepackage{showlabels}
\usepackage{graphicx}
\usepackage{epstopdf}
\usepackage[utf8]{inputenc}
\usepackage{enumerate}
\usepackage{cite}
\usepackage{caption}
\usepackage{subcaption}
\usepackage{hyperref}
\usepackage{braket}
\usepackage{titlesec}
\usepackage{fixltx2e}
\usepackage{cleveref}
\hypersetup{breaklinks}
\usepackage[usenames,dvipsnames,svgnames,table]{xcolor}
\usepackage{comment}
\newlength\mytemplen
\newsavebox\mytempbox

\makeatletter
\newcommand\mybluebox{%
    \@ifnextchar[%]
       {\@mybluebox}%
       {\@mybluebox[0pt]}}

\def\@mybluebox[#1]{%
    \@ifnextchar[%]
       {\@@mybluebox[#1]}%
       {\@@mybluebox[#1][0pt]}}

\def\@@mybluebox[#1][#2]#3{
    \sbox\mytempbox{#3}%
    \mytemplen\ht\mytempbox
    \advance\mytemplen #1\relax
    \ht\mytempbox\mytemplen
    \mytemplen\dp\mytempbox
    \advance\mytemplen #2\relax
    \dp\mytempbox\mytemplen
    {\hspace{1em}\usebox{\mytempbox}\hspace{1em}}}

\makeatother
\def\c{\cite}
\def\so{\mathfrak{so}}
\def\su{\mathfrak{su}}
\def\u{\mathfrak{u}}
\def\foot{\footnote}
\def\M{{\cal M}}
   \def\Rc{{\cal R}}
   
   \def\Pc{{\cal P}}
   \def\dint{\int\!\!\!\int}

\def\P{\Phi}

\def\DP{\Delta_{\Phi}}

\def\o{\over}

\def\d{\delta}

\def\0{{(0)}}
\def\1{{(1)}}
\def\2{{(2)}}
\def\3{{(3)}}
\def\4{{(4)}}

\def\G{\Gamma}

\def\gap{\text{gap}}

\def\eqr{\eqref}
\def\sec{\section}

\def\l{\lambda}
\def\vs{\vskip .1 in}

\def\a{\alpha}
\def\rar{\rightarrow}

\def\la{\langle}
\def\ra{\rangle}
\def\O{{\cal O}}

\def\ssec{\subsection}
\def\sssec{\subsubsection}
\def\sec{\section}
\newcommand{\Farg}[3]{\left( \begin{array}{c} #1 \\ #2\end{array} \Big| #3 \right)}
\def\i{\infty}

\def\F{{\cal F}}
\def\O{{\cal O}}
\def\ra{\rangle}
\def\la{\langle}

\def\t{\tau}

\def\s{\sigma}

\def\vs{\vskip .1 in}
\def\D{\Delta}

\def\g{\gamma}
\def\a{\alpha}
\newcommand {\be} {\begin {equation}}
\newcommand {\ee} {\end {equation}}

\newcommand {\bes} {\begin {equation*}}
\newcommand {\ees} {\end {equation*}}

\newcommand{\es}[2] {\begin{equation} \label{#1} \begin{split} #2 \end{split} \end{equation}}
\newcommand{\e}[2] {\begin{equation} \label{#1} #2 \end{equation}}

\newcommand{\Z}{\mathbb{Z}}
\newcommand{\N}{\mathbb{N}}
\newcommand{\R}{\mathbb{R}}

\newcommand{\beq}{\begin{equation}}
\newcommand{\eeq}{\end{equation}}

\newcommand{\p}{\partial}

\def\be{ \begin{equation} }
\def\ee{ \end{equation} }
\def\N{{\cal N}}

\def\eqr{\eqref}

\def\b{\beta}
\def\l{\lambda}

\def\rar{\rightarrow}

\def\eps{\epsilon}

 \newmuskip\pFqmuskip

\newcommand*\pFq[6][8]{%
  \begingroup % only local assignments
  \pFqmuskip=#1mu\relax
  \mathcode`\,=\string"8000
  \begingroup\lccode`\~=`\,
  \lowercase{\endgroup\let~}\pFqcomma
  {}_{#2}F_{#3}{\left[\genfrac..{0pt}{}{#4}{#5};#6\right]}%
  \endgroup
}
\newcommand{\pFqcomma}{\mskip\pFqmuskip}
\makeatletter
\renewcommand{\@maketitle}{
\newpage
 \begin{center}%
  {\large\bfseries \@title \par}%
 \end{center}%
 \par} \makeatother
\numberwithin{equation}{section}

\begin{document}
\preprint{PUPT-2536\\ CALT-TH 2017-054}

\institution{PU}{Department of Physics, Princeton University, Princeton, NJ, 08544, USA}
\institution{CT}{Walter Burke Institute for Theoretical Physics, California Institute of Technology, \cr Pasadena, CA, 91125}

%\title{Draining the Swampland \\with Double-Trace Flows}
\title{Double-Trace Flows and the Swampland}

\authors{Simone Giombi\worksat{\PU}, Eric Perlmutter\worksat{\PU, \CT}}

\abstract{We explore the idea that large $N$, non-supersymmetric conformal field theories with a parametrically large gap to higher spin single-trace operators may be obtained as infrared fixed points of relevant double-trace deformations of superconformal field theories. After recalling the AdS interpretation and some potential pathologies of such flows, we introduce a concrete example that appears to avoid them: the ABJM theory at finite $k$, deformed by $\int\!\O^2$, where $\O$ is the superconformal primary in the stress-tensor multiplet. We address its relation to recent conjectures based on weak gravity bounds, and discuss the prospects for a wider class of similarly viable flows. Next, we proceed to analyze the spectrum and correlation functions of the putative IR CFT, to leading non-trivial order in $1/N$. This includes analytic computations of the change under double-trace flow of connected four-point functions of ABJM superconformal primaries; and of the IR anomalous dimensions of infinite classes of double-trace composite operators. These would be the first analytic results for anomalous dimensions of finite-spin composite operators in any large $N$ CFT$_3$ with an Einstein gravity dual.}

\date{ }

\maketitle
\setcounter{tocdepth}{3}
\tableofcontents

\sec{Introduction}

The paradigmatic example of AdS/CFT \c{mald, gubs, witt} is a duality between two theories: a theory of AdS quantum gravity whose low-energy limit includes a weakly coupled Einstein gravity subsector; and a CFT with many degrees of freedom (``large $N$''), and a sparse spectrum of local, single-trace operators of spin no greater than two. The absence of parametrically light single-trace operators of higher spin is often phrased as a holographic gap condition, $\D_\gap\gg 1$, where $\D_\gap$ is the dimension of the lightest spin-four single-trace operator \c{Heemskerk:2009pn}. As a matter of consistency, these conformal field theories must, among other properties, be unitary, crossing-symmetric and causal; must they also be supersymmetric? 

There are recent conjectures in the affirmative, motivated in part by an absence of explicit constructions. It was argued in \c{OV, FK} that weak gravity bounds on the tension of charged objects in quantum gravity may only be saturated when a BPS condition is obeyed; combined with known instabilities of non-BPS branes in AdS spacetimes, \c{OV, FK} argued that, indeed, sparse non-supersymmetric CFTs with large $N$ and large gap do not exist. This connection arises because the weak gravity bounds are meant to hold in theories whose low-energy gravitational sector is described by general relativity, a feature which is automatically implied by the CFT spectral condition $\D_\gap\gg1$ \c{cemz, hartman, chp}.

In this paper, we explore a general approach to the construction of large $N$, large gap, non-supersymmetric CFTs, and investigate specific examples that avoid some well-known pathologies that are both perturbative and non-perturbative in $N$. The main idea is to impose supersymmetry-breaking boundary conditions in AdS. On the CFT side, we perturb a UV superconformal field theory by a relevant, supersymmetry-breaking, double-trace deformation, which flows to an IR CFT that still obeys the gap condition. The construction is on firmest footing when the UV SCFT has no symmetry-preserving exactly marginal operators. 
While not all such flows may lead to a viable IR fixed point, 
we present a large class of examples in three spacetime dimensions that appear to be especially promising.\foot{There is a long history of attempts at consistent AdS/CFT constructions without supersymmetry. Some are cited in \c{OV}. Those with a direct connection to M2-branes include \c{dnp, distler, berkooz, murugan, naqvi, Fischbacher:2010ec, Gaiotto:2009mv, banks}.} This construction may be viewed as a potential counterexample to the CFT conjecture of \c{OV, FK}, but is consistent with their sharpened version of the weak gravity bounds as applied to flat space quantum gravity.

%%%%%%%%%%%
%%%%%%%%%%%%%%%

In Section \ref{s2}, we introduce the basic idea of a supersymmetry-breaking double-trace flow from a large $N$, large gap SCFT; make a connection to the conjectures of \c{OV, FK}; list some pathologies that a consistent construction must avoid; and discuss the most basic constraints on such flows from superconformal representation theory in various spacetime dimensions. 

In Section \ref{s3}, we introduce a specific class of examples: namely, the ABJM theories \c{ABJM} at finite $k$ deformed by $\int\!\O^2$, where $\O$ is the superconformal primary in the stress-tensor multiplet. We argue for their viability at large but finite $N$, and address a potential issue on the moduli space of vacua. We also discuss a preliminary proposal for using double-trace deformations of more general three-dimensional CFTs to generate other non-supersymmetric theories with a large gap. 

In Section \ref{s4}, having made a concrete proposal for a non-supersymmetric, large gap CFT, we perform some quantitative computations of its observables. We show that certain classes of correlation functions vanish in the IR at {\it leading} order in $1/N$, and very briefly discuss the implications for thermal physics. We then study the change under RG flow of an infinite class of four-point functions of ABJM superconformal primaries $\O_p$. This calculation generalizes the approach of \c{kirilin}. By performing the conformal block decomposition of those results, we extract, in Section \ref{4.3}, the leading-order change in the anomalous dimensions of spinning double-trace operators $\O_p \p_{\mu_1}\ldots\p_{\mu_\ell}\O_p$. Moreover, using the non-renormalization properties of UV-protected supermultiplets, we can extract the anomalous dimension of several classes of double-trace operators {\it at the IR fixed point}, not only their change under RG flow. The main results may be found in \eqr{chan}, \eqr{ad1}--\eqr{ad2}, and \eqr{prot}. These would be, to our knowledge, the first analytic computations of anomalous dimensions of finite-spin double-trace operators in any large $N$, three-dimensional CFT with an Einstein gravity dual, supersymmetric or otherwise. Our technique for analytically deriving these anomalous dimensions applies to any IR fixed point obtained by double-trace flow from a CFT with a greater number of supersymmetries, including, in particular, IR SCFTs that preserve a fraction of the UV supersymmetry. 

In Section \ref{s5}, we conclude with some comments on future work. Appendices \ref{appa}--\ref{appb} include some technical details needed for Section \ref{s4}, while Appendix \ref{appc} gives a brief historical recollection of the attempt to construct non-supersymmetric, large gap CFTs by orbifolding $\N=4$ super-Yang-Mills.

Before proceeding, it is worth being more precise about our definitions, and our interpretation of the conjectures of \c{OV, FK} that motivated this work. In general, a CFT with a weakly coupled gravity dual belongs to a sequence of CFTs with central charge $C_T$ parameterized by $N$. The sequence exists above some critical value of $N$ and admits a $N\rar\i$ limit with a finite spectral density and operator product expansion. The prototypical holographic CFT, whose gravity sector is general relativity, obeys the further constraint $\D_\gap\gg 1$, and the ``sparseness'' condition of polynomially-bounded growth of low-spin operators. (This can include towers of KK modes from transverse manifolds with AdS-scale curvatures.) It is this class of CFTs whose existence we are addressing. It should be emphasized that the question of whether holographic CFTs can have a parametrically large gap ceases to make sense at small enough $N$: these theories always obey a parametric hierarchy $\D_\gap \lesssim \sqrt{C_T}$. (This was recently proven in CFT in \c{caron-huot}.) In string theory terms, $M_{string} \lesssim M_{Planck}$. For finite $N$ and $\D_\gap$, UV-finiteness demands corrections to the gravitational action beyond general relativity, at which point it is unclear what becomes of the conjecture of \c{OV, FK} then.

\sec{Basic Idea}

Let us first quickly recall the definition of a double-trace flow. Consider a large $N$ CFT$_d$ which contains a scalar conformal primary $O$ of conformal dimension $\D<d/2$. If we deform the action by
\e{dt}{\delta S_{\rm CFT}=g \int d^d x \,O ^2~,}
this triggers a flow to an IR CFT, in which $\D\rar d-\Delta+\ldots$ to leading order in $1/N$. If $O$ is not a singlet under global symmetries, $O^2$ is to be understood as the singlet in the operator product. The generating functional for connected correlators of the IR CFT is given by the Legendre transform of the UV functional with respect to a source for $O$ \cite{kw2}. At infinite $N$, such flows always exist. Besides the change $\D\rar d-\D$, the spectrum of the planar dilatation operator is identical in the UV and IR. At the non-planar level, conformal dimensions and OPE coefficients are modified, both for single-trace and multi-trace operators. The change in the double-trace spectrum was recently analyzed in \c{kirilin}, and we will reprise those results later in this paper.

 \begin{figure}
    \centering
       \includegraphics[width = .6\textwidth]{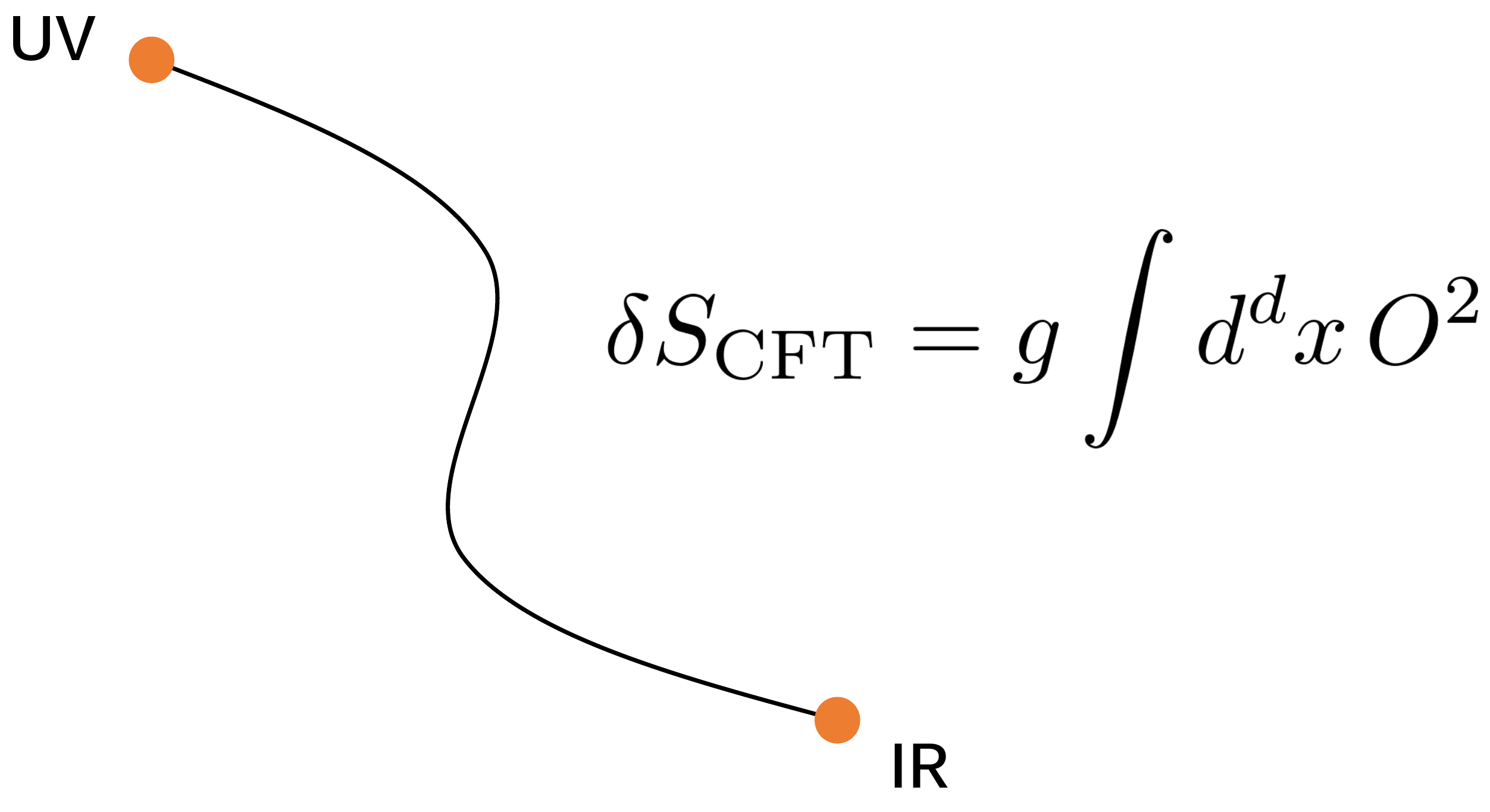}
         \caption{A cartoon of a general double-trace flow. In this paper, we take the UV theory to be a superconformal field theory with large $N$ and a large gap, and $O$ to be a superconformal primary.}
         \label{fig4}
\end{figure}

In the holographic context \c{kw2, wittendt, Mueck:2002gm, Berkooz:2002ug, Gubser:2002zh, Gubser:2002vv, HartmanRastelli, Diaz:2007an}, in which $O$ is dual to a scalar field $\phi$ of mass squared $m^2 = \D(\D-d)$ in AdS units, there are two choices of normalizable boundary conditions when $-{d^2/4} \leq m^2\leq -{d^2/4}+1$: 
\es{}{\text{\bf Standard:}\quad &\D_+ = {d\o 2}+\sqrt{{d^2\o 4}+m^2}\\
\text{\bf Alternate:}\quad &\D_- = d-\D_+}
Each of these corresponds to a unitary conformal dimension at one end of the RG flow triggered by \eqr{dt}. The flow is not visible in the bulk as a soliton interpolating between two AdS vacua with a macroscopic difference in curvatures, due to $1/N$ suppression. 

We now consider this deformation in a top-down setting. Consider a SUSY AdS$_{d+1}\times \M$ compactification of string or M-theory. In the limit in which the bulk theory is a weakly coupled supergravity and $\M$ is of AdS size, as in Freund-Rubin compactifications, the dual SCFT has a parametrically large gap to single-trace operators of spin greater than two, and the light spectrum is sparse.\foot{In $d=2$, a less restrictive and more explicit definition of sparseness, $\rho(\D) \lesssim e^{2\pi \D}$ for $\D-\D_{\rm vac}\leq c/12$, is sufficient to capture many aspects of holographic universality \c{hks}.} If the Kaluza-Klein spectrum on $\M$ contains scalar fields with masses in the aforementioned range, SUSY may require alternate quantization. This happens, for instance, for bottom components of chiral multiplets of unit $R$-charge in 4d $\N=1$ SCFTs (where $\D=3/2$), or of flavor current multiplets in 3d $\N=2$ SCFTs (where $\D=1$). In such cases, we may deform the dual SCFT as in \eqr{dt}, where $O$ is the conformal primary operator dual to such a scalar field. Such a double-trace deformation will generically break all SUSY in the IR. Thus, by choosing the SUSY-breaking $\D_+$ boundary condition in AdS, we are studying a non-SUSY, large $N$ fixed point with $\D_\gap\gg1$. 

Let us make contact with \c{OV, FK}. Their proposed sharpening of the weak gravity conjecture states that in a consistent theory of quantum gravity, the weak gravity bounds on charge-to-mass ratios may only be saturated by BPS objects. AdS spacetimes suffer from a brane nucleation instability for non-BPS branes with a sufficiently large charge-to-tension ratio\c{mms, switt}, which destroys the near-horizon region of a stack of branes in asymptotically (locally) flat space; \c{OV, FK} take this as evidence that non-SUSY CFTs with large $N$ and a large gap do not exist. But the special class of non-SUSY CFTs studied herein -- obtained by SUSY-breaking boundary conditions in AdS geometries built from BPS branes -- may nevertheless be consistent with the refined bound on charged objects in flat space. In the brane picture -- that is, before taking the near-horizon limit -- there is no boundary, and hence no notion of SUSY-breaking boundary condition.\foot{The sharpened weak gravity bounds are implied by \c{OV, FK} to hold in both flat space and AdS quantum gravity. The natural argument for the latter, assuming the former, is that AdS geometries in string/M-theory are canonically constructed by bringing BPS branes together and zooming into the near-horizon region of the backreacted geometry. But logically speaking, these seem to be distinct claims. We thank H. Ooguri for a discussion on this issue. See also \cite{Nakayama:2015hga} for a discussion of the weak gravity bounds in AdS. } We return to this point in Section \ref{s3}.

\sssec*{What can go wrong?}

Obviously, the question is how robust this construction is away from infinite $N$. But some pathologies are absent by design. At infinite $N$ the CFT has a unitary spectrum of operator dimensions; in the bulk, the classical theory is perturbatively stable. Moreover, non-perturbative instabilities arising from geometry or topology of $\M$ are also absent. 

There are, still, various potential pitfalls that may render the IR CFT ill-defined, including {\bf i)} complex operator dimensions in the $1/N$ expansion; {\bf ii)} an unstable vacuum; {\bf iii)} the breaking of conformal symmetry via nonzero beta functions for marginal or nearly-marginal gauge-invariant operators, which can be generated for single-trace or multi-trace operators. If the CFT has a conformal manifold, the offending pathology -- and its bulk dual -- may change as a function of marginal couplings (as in e.g. \c{adams, hop}). 

On the bulk side, one may likewise find {\bf i)} bulk tachyons that violate the Breitenlohner-Freedman bound \c{bf}. This can occur at tree- or loop-level, but the most dangerous situation is when a tree-level scalar mass sits at the bound, $m^2_{BF} = -{d^2/4}$: in the absence of SUSY, quantum corrections may push $m^2< m^2_{BF}$; {\bf ii)} the development of a runaway direction in the effective potential for probe branes in AdS; {\bf iii)} non-perturbative instabilities due to the geometry and/or topology of $\M$; {\bf iv)} non-perturbative instabilities in AdS, such as the potential for brane nucleation of \c{mms, switt}. 

\sssec*{In which spacetime dimensions?}\label{s2}
With these issues in mind, let us briefly comment on the viability of this construction in various space-time dimensions, before moving on to the case of $d=3$, our main interest.

In $d=5,6$, superconformal representation theory prohibits the existence of (unitary) scalar conformal primaries in the range ${d-2\o 2}<\D<{d\o 2}$ (see e.g. \c{cordova}).\footnote{This eliminates one approach to constructing the elusive, and perhaps non-existent, interacting non-SUSY $d>4$ CFT.} In $d=4$, the same is true for maximal $\N=4$ SUSY, but $\N<4$ theories can (and do) contain operators with $1<\D<2$. In $d=2,3$, the superconformal algebra admits unitary representations containing scalar primaries of $\D<d/2$ for any amount of SUSY. 

In $d=4$ SCFTs, a ubiquitous class of operators with $\D<2$ are $\N=1$ chiral primaries with unit $\u(1)_R$-charge, which have $\D={3/2}$. However, finding viable examples of SUSY-breaking double-trace flows to stable IR fixed points in $d=4$ seems challenging. The bottom component of a four-dimensional $\N=1$ current multiplet is a $\D=2$ superconformal primary, $\O$. The dual scalar field in AdS sits exactly at the BF bound. One can also form double-trace operators $\O^2$ that are classically marginal global symmetry singlets; such operators are typically marginally relevant, and may develop nonzero beta functions along a conformal manifold. The latter issue also applies to composites made from fermions in $\N=1$ chiral and anti-chiral multiplets, which have $\D=2$. An explicit example of a large $N$, large gap CFT where these problems would arise is the Klebanov-Witten theory, dual to type IIB string theory on AdS$_5\times T^{1,1}$ \cite{KW}. The theory contains an operator of unit $R$-charge, $\O_1$, that lives in the $(2,2)$ of an $\su(2)_A\times \su(2)_B$ global symmetry. This is the only operator of $\D<2$. Thus, if we turn on the deformation
\e{}{\delta S_{KW}  = g \int (\O_1)_{a b}(\overline\O_1)^{a b}~,}
by choosing the $\D_+$ boundary condition on the dual bulk scalar field, the CFT will naively flow to a non-SUSY fixed point. This was briefly considered in \c{strassler, gk, Gubser:2002zh}. But for the reasons given above, the putative IR fixed point is unlikely to exist. (The Klebanov-Witten theory also has an exactly marginal coupling, which can acquire a nonzero beta function in the $1/N$ expansion after double-trace flow.)

On the other hand, the representation theory of $\mathfrak{osp}(\N|4)$, the $d=3$ superconformal 
algebra, is much more favorable, and we focus on this case below.   

\sec{A SUSY-Breaking Double-Trace Flow from ABJM}\label{s3}
We now turn to our main proposal: that a non-SUSY double-trace flow from the ABJM theory leads to a stable IR fixed point. 

Let us quickly review the salient aspects of the ABJM theory that we will need. (For more detailed reviews, see e.g. \c{torri, lambert, pufuf}.) The ABJM theory \c{ABJM} is a parity-invariant, $U(N)_k\times U(N)_{-k}$ Chern-Simons theory coupled to bifundamental matter. For $k=1,2$, the theory has $\N=8$ SUSY, with $\mathfrak{so}(8)$ $R$-symmetry, whereas for $k>2$ it has $\N=6$ SUSY, with $\su(4)_R\times \u(1)\subset \mathfrak{so}(8)$ $R$-symmetry. The nature of the large $N$ limit depends on whether $k$ is held finite. At finite $k$, the holographic dual is 11d supergravity on AdS$_4\times S^7/\Z_k$, supported by four-form flux. At large $k$, one may take a `t Hooft limit, $k\rar\i, N\rar \i$, with $\l=N/k$ fixed, where, up to numerical prefactors, $\a'\sqrt{\l}\sim 1$. Upon increasing $k$ while keeping $\l\gg1$, the bulk dual remains 11d supergravity up to $k^5\sim N$, where it crosses over to type IIA supergravity on AdS$_4\times \mathbb{CP}^3$. Eventually, for large enough $k$, stringy effects become parametrically important. It is useful to view the $S^7/\Z_k$ as a circle fibered over $\mathbb{CP}^3$ with length $2\pi L_{\rm AdS}/k$; this essentially corresponds to the $\u(1)$ symmetry of the CFT. The ``central charge'' $C_T$ of the ABJM theory is
\e{}{C_T \approx {64\o 3\pi} \sqrt{2k}N^{3/2} + \O(\sqrt{N})}
where $C_T$ is defined by the stress tensor two-point function as
\e{}{ \la T_{\mu\nu}(x)T_{\rho\sigma}(0) \ra = C_T {{\cal I_{\mu\nu\rho\sigma}}(x)\o x^{6}}}
with a fixed tensor structure ${\cal I_{\mu\nu\rho\sigma}}(x)$ defined in \cite{petkou}. 

For any value of $k$, the bottom component of the stress tensor multiplet is a $\D=1$ scalar, which we call $\O$. For $k=1,2$, $\O$ resides in the ${\bf 35_c}$ of the $\mathfrak{so}(8)$ $R$-symmetry,\footnote{We have made a choice of $\mathfrak{so}(8)$ triality frame, following conventions of \c{pufu}. See e.g. Table \ref{tab1}.} with Dynkin labels [0020]; we may represent it as $\O_{IJ}$, the symmetric, traceless rank-two tensor of $\mathfrak{so}(8)$, where $I,J=1\ldots 8$. In terms of the fundamental scalars $X_I$ in the ABJM Lagrangian, 
\e{}{\O_{IJ} = \text{tr}(X_{I}X_{J} - {\d_{IJ}\o 8} X^2)}
For $k>2$, $\O$ resides in the ${\bf 15}$ of the $\su(4)$ $R$-symmetry, with $\su(4)$ Dynkin labels [101]; we may represent it as $\O^a_b$. In either case, from these components, one may form many double-trace operators, preserving varying degrees of $R$-symmetry. A natural choice is to preserve the full $R$-symmetry, but break all SUSY. Thus, we propose to consider the RG flow away from the ABJM theory triggered by the $R$-symmetry singlet $\O^2$, with all $R$-symmetry indices contracted:
\e{rg}{\delta S_{ABJM} = g\int d^3 x\, \O^2}
where 
\es{}{\O^2&=\O_{IJ}\O^{IJ} ~~\text{for}~~ k=1,2\\
\O^2 &= \O^a_b\O_a^b ~~\,~~\text{for}~~k>2~.}
This relevant deformation triggers a flow to an IR fixed point where ${\cal O}$ has dimension 
$\Delta_{\cal O}=2+O(1/N)$. In the dual AdS theory, this corresponds to choosing $\Delta_+$ boundary conditions for all components 
of the $m^2=-2$ scalar field dual to ${\cal O}$. Our proposal is that such IR fixed points may be viable CFTs for finite $k$. More precisely, this statement is on firmest footing for $k\neq 1,3$, as we explain below. 

\vs
Let us lay out some arguments for this. First, as mentioned earlier, the AdS$_4\times S^7/\Z_k$ geometry admits no tunneling solutions that can be ascribed to the transverse manifold, because it descends from a SUSY compactification. Second, at finite $k$ the ABJM theories form a discrete set, with no exactly marginal gauge coupling. This is one virtue of AdS$_4\times \M_7$ compactifications in general as compared to AdS$_5\times \M_5$ compactifications.

Another appealing property of the finite $k$ ABJM theories subject to the $R$-symmetry-preserving double-trace deformation \eqr{rg} is their especially sparse spectrum of light operators:  all symmetry-preserving gauge-invariant operators in the IR are irrelevant. 
More precisely, the IR fixed point has the following two properties, to all orders in $1/N$: 
\vs
{\bf A~} There are no global singlet, parity-preserving relevant operators. 
\vs
{\bf B} \,\,\,There are no global singlet, parity-preserving marginal operators for $k\neq 1,3$. 

\vs
\noindent {\bf B} implies that there are no $\D={3/2}$ scalar operators, which would be dual to BF bound-saturating scalars. (If there were, we could use them to form a classically marginal singlet double-trace operator.) Together these imply the complete attractiveness of the double-trace RG flow and the absence of conformal symmetry breaking.

To show {\bf A} and {\bf B}, we examine the spectrum of gauge-invariant operators of the ABJM theory. The spectrum at large $N$ may be obtained by starting from the KK reduction on $S^7$ \c{Castellani:1984vv,Duff:1986hr}, and restricting to $\Z_k$-invariant modes. See \c{liu} for a summary. These are the modes with $\u(1)$ charge 0 mod $k$.\footnote{For instance, the fact that $\O$ lives in the adjoint of $\su(4)_R$ in the $k>2$ ABJM theories may be easily understood via the projection onto $\Z_k$-invariant states, together with the branching of the $35_c$ of $\mathfrak{so}(8)$ into $\su(4)\times \u(1)$ representations,
\e{}{{\bf 35_c} \rar {\bf 15}_0 + {\bf10}_2 + {\bf10}_{-2}}
where the subscript labels the $\u(1)$ charge. Henceforth we will show only $\mathfrak{so}(8)$ Dynkin data $[abcd]$ explicitly, leaving the branching into $\su(4)\times \u(1)$ for the $\N=6$ theories implicit.} The scalar spectrum on $S^7$ is comprised of a tower of KK modes in $\mathfrak{so}(8)$ representations $[00p0]$, where $p=2,3,\ldots$. The dual operators are single-trace, superconformal primaries which we call $\O_p$, with $\D_p={p/2}$. Each of these is a bottom component of a 1/2-BPS superconformal multiplet, namely, the $B_1[0]_{p/2}^{(0,0,p,0)}$ multiplets in the notation of \c{cordova}. The $p=2$ multiplet is the stress tensor multiplet, which contains $\O_2\equiv \O$; we list its content in Table \ref{tab1}. Under the branching $\so(8) \rar \mathfrak{su}(4)\times \mathfrak{u}(1)$, the representation $[00p0]$ yields $\u(1)$ charges $\pm (p-2n)$ with $n=0,1,\ldots \lfloor{p/ 2}\rfloor$.

\begin{table}%[htdp]
\begin{center}
 \begin{tabular}{|c|c|c|l|}
\hline
Operator& $\D$ & $\ell$ & ~~~~~{$\mathfrak{so}(8)$}  \\
  \hline
 $\O$ &$1$ & $0$ & ${\bf 35}_c = [0020]$ \\
$Q\O$&  $3/2$ & $1/2$ & ${\bf 56}_v = [0011]$ \\
$Q^2\O$&  $2$ & $0$ & ${\bf 35}_s = [0002]$\\
$Q^2\O$&  $2$ & $1$ & ${\bf 28} = [0100]$\\
$Q^3\O$&  $5/2$ & $3/2$ & ${\bf 8}_v = [1000]$ \\
$Q^4\O$&  $3$ & $2$ & ${\bf 1} = [0000]$\\
\hline
 \end{tabular}
\end{center}
\caption{The conformal primaries of the stress-tensor multiplet in a 3d, $\N=8$ SCFT, together with their $\so(3,2)\times \so(8)$ quantum numbers and their positions in the multiplet. This is the $B_1[0]_1^{(0,0,2,0)}$ multiplet in the notation of of \c{cordova}. The superconformal primary is $\O$. In a theory with $\N=6$ SUSY, the first three columns are identical.}
\label{tab1}
\end{table}%

We now look to construct relevant and marginal scalar operators at the IR fixed point that are singlets under all global symmetries. %: it is this class of operators that could in principle acquire nonzero beta functions in the $1/N$ expansion. 
We have broken SUSY completely, but kept the full $R$-symmetry. The theory is also parity-invariant. So we are looking for parity-even $R$-singlets. As described above, all single-trace scalars are charged under the $R$-symmetry, so only multi-trace operators can be $R$-singlets. In the UV, before the RG flow, the only operators that can possibly be used to form marginal or relevant multi-trace scalar singlets have $\D=1,{3/2},2$. Taking $k\neq 1,3$, which removes the potentially problematic $\O_3$ from the spectrum, the list of such operators is short:

\begin{itemize}
\item  $\O$, with $\D=1$ in the ${\bf 35_c}$, and its superconformal descendants with $\D\leq 2$. These include spin-1/2 fermions $\psi$, with $\D=3/2$ in the ${\bf 56_v}$, and scalars, with $\D=2$ in the ${\bf 35_s}$ -- see Table \ref{tab1}. 

\item $\O_4$, with $\D=2$ in the ${\bf 294_c}=[0040]$. 
\end{itemize}

\noindent Due to $\so(8)$ selection rules, the only relevant singlet comprised of these constituents is $\O^2$, our deforming operator. In 
the UV, there is also a nearly marginal triple-trace operator obtained from $\O^3$ (projecting onto the singlet part). However, the RG flow \eqr{rg} takes $\D \rar 2+O(N^{-3/2})$, for {\it all} components of $\O$. This immediately implies that for generic $k$, there are no relevant singlets in the IR.\footnote{
We do not expect that the presence of the nearly marginal triple-trace operator $\O^3$ in the UV renders the RG flow \eqr{rg} invalid. A similar situation 
arises in the standard Wilson-Fisher fixed point in the 3d $O(N)$ model: there is a nearly marginal ``triple-trace" operator $(\phi^i\phi^i)^3$ 
in the UV, but this does not affect the RG flow triggered by the relevant operator $(\phi^i\phi^i)^2$. See also \c{GY} for a similar example with SUSY in the UV.} Likewise, the only candidate (nearly-)marginal singlets in the IR are the two-fermion operators, :$\psi\psi$:, projected onto the singlet. %\footnote{For $k=2$ for instance, the singlet is unique.}
However, this operator is parity odd.\foot{This can be seen by noting that in any CFT with a weakly coupled AdS dual, the $1/N$ expansion may be viewed as an expansion around generalized free fields. The parity of a given operator is discrete, so we may determine it at infinite $N$. It is known that, in the theory of  generalized free spin-1/2 fermions, the scalar operator :$\psi\psi$: with dimension $2\D_\psi$ is parity odd. This can be seen, for instance, by decomposing the four-point function of identical, generalized free spin-1/2 fermions.} Thus, we have shown Properties {\bf A} and {\bf B}. These imply that the RG flow \eqr{rg} leads to a stable 
fixed point, and the IR CFT admits no symmetry-preserving relevant flows. 

\vs
Having established the absence of relevant and marginal global singlet operators, let us however note that such operators 
would not have been expected to spoil the existence of the IR fixed point anyway. 
Given some set of operators $\O_i$ with $\D_i-d=\eps_i\ll1$ and corresponding couplings $g^i$, ordinary conformal perturbation theory admits both trivial $(g^i=0)$ and non-trivial ($g^i\sim \eps^i$) fixed points. For either sign of $\eps^i$, the trivial fixed point is guaranteed to exist. This should be contrasted with the case of non-SUSY orbifolds of $\N=4$ SYM in the `t Hooft limit, in which, due to the exactly marginal `t Hooft coupling, nearly-marginal global singlets develop nonzero beta functions to any order in perturbation theory \c{dym3, dym2, rastelli}. %Indeed, our restriction to finite $k$ avoids a potential problem in the `t Hooft limit of the ABJM theory: the marginal coupling $\l$ may acquire a nonzero beta function in the $1/N$ expansion. 
Note, though, that even when there are flat directions in a theory with nearly-marginal singlets, the addition of a {\it relevant} deformation, like $\int \O^2$, may still lead to a stable IR fixed point. This happens in e.g. \c{9904017, KW, klose}; while these examples retain some SUSY in the IR, perhaps the same also happens in the `t Hooft regime of the ABJM theories after our SUSY-breaking double-trace deformation.

\ssec{A general prescription for 3d CFT}
There are obvious variations of the above, including extension to the $\N=6$ $U(N)_k\times U(M)_{-k}$ ABJ theories \c{ABJ}, or the introduction of double-trace deformations that preserve only a subgroup of $\so(8)$. More interesting are generalizations beyond ABJ(M). Consider an AdS$_4\times \M_7$ solution of 11d supergravity, where $\M_7$ is a Sasaki-Einstein manifold. The dual $\N=2$ SCFT is, like ABJM, isolated. A preliminary proposal for generating stable, IR non-SUSY CFTs is to change the AdS boundary conditions for {\it all} single-trace scalar operators with $\D<3/2$. That is, for every such operator $\O_i$ with $\D_i<3/2$, deform the UV SCFT action by
\e{}{\d S_{CFT} = \sum_{\lbrace\O_i|\D_i<{3\o 2}\rbrace} g^i \int d^3x \,\O_i^2}
for some couplings $g^i$. Following our ABJM discussion, the question of IR stability boils down (modulo issues we have not exorcised yet) to the question of whether there are any single-trace scalar operators with $\D=3/2$. 

Examples abound. In $d=3$, the bottom component of $\N=2$ flavor multiplets is a $\D=1$ scalar. %This includes the $R$-current multiplet, so every $\N=2$ SCFT admits such a double-trace flow. 
($\N=1$ flavor multiplets do not contain scalars, so $\N=1$ SCFTs may, but need not, contain a $\D<3/2$ scalar with which to flow.) A well-studied $\N=2$ example is the 11d supergravity compactification on AdS$_4\times M^{111}$ \c{fabbri, fabbri2}, dual to a quiver gauge theory with three nodes \c{Martelli:2008si}. This manifold has isometry group $G = \su(3)\times \su(2)\times \u(1)_R\times \u(1)_B$, where the $\u(1)_B$ is a baryonic or ``Betti'' vector multiplet. From the analysis of the KK spectrum in \c{fabbri} (see also Table 5.2 of \c{murugan}), one can check that the CFT contains no $\D=3/2$ single-trace scalar operators.\foot{Any manifold $\M_7$ with topologically non-trivial two-cycles has $b_2$ Betti multiplets. The dual CFT has a global symmetry group $G=G'\times \u(1)^{b_2}$. Members of Betti multiplets are $G'$-singlets. The top component of a Betti multiplet is a $\D=2$ scalar. Thus, in CFT$_3$'s with Betti multiplets, the putative IR fixed point obtained after double-trace flow has (at least) $b_2$ relevant, single-trace, parity-even $G$-singlet scalars. Accordingly, while the IR fixed point may be stable, it is thus not a ``dead-end'' CFT.} It would be worthwhile to examine this proposal further.

\ssec{Comments}
\sssec*{Connection to other proposals}
The double-trace technique considered in this paper is ``milder'' than other SUSY-breaking constructions that break SUSY in the bulk, not just by boundary conditions. In Appendix \ref{appc}, we briefly recall the story of one of the most well-studied -- and ultimately unsatisfactory -- non-SUSY constructions, namely, the type IIB orbifolds of the form AdS$_5\times S^5/\Gamma$, dual to non-SUSY orbifolds of 4d $\N=4$ SYM.

In \c{banks}, a morally similar construction (inspired in part by \c{Dong1, Dong2}) was suggested for the $k=1$ ABJM theory, in which the $\so(8)$ $R$-symmetry is gauged and augmented by a Chern-Simons term. This may be implemented by a double-trace deformation $\d S_{\rm CFT} = \int d^3 x J_\mu J^\mu$, where $J_{\mu}$ is the $R$-symmetry current, which is induced holographically through a mixed boundary condition on the bulk gauge field \c{witten2003}. Though the IR fixed point may indeed exist, it has more potentially problematic operators whose dynamics may destabilize the theory and/or drive it to non-unitarity (such as the triple-trace singlet $\O^3$ discussed in \c{banks}); it also would have the relevant singlets $\O^2$, thus making the RG flow less stable. Our proposal is simpler, and eliminates these operators, as described above.

\sssec*{Moduli space of vacua}
A potential issue with this class of theories is the stability of the moduli space of vacua. In the ABJM theory on $\R^3$, the effective potential on the moduli space vanishes. In the IR, in the absence of SUSY, these flat directions will presumably be lifted at finite $N$; in principle, the origin of moduli space could cease to be a minimum, or runaway instabilities could develop in the $1/N$ expansion.\foot{See \c{tseytlin, adams, ale} for discussions of similar instabilities in the non-SUSY $\N=4$ SYM orbifold context. In \c{rastelli}, it was shown that in fixed lines of 4d CFTs with adjoint matter, Coleman-Weinberg instabilities exist if and only if conformal symmetry is broken via nonzero beta functions. If this equivalence extends to the present case (though we know no reason this would be so), our previous arguments about the spectrum imply an absence of moduli space instabilities for the ABJM double-trace deformation.} Definitively understanding the fate of the moduli space appears to be highly involved, but let us make some observations.\foot{We thank O. Aharony for raising this issue and for valuable discussions about it.} 

Intuitively, the deformation \eqr{rg} appears to lift the moduli space. With $g>0$, it is a stabilizing quartic potential for all ABJM scalars, that grows as one flows toward the IR. It is nevertheless possible that, in the deep IR, the minimum is pushed away from the origin, or worse, by higher order effects in $1/N$. (If there is indeed a minimum away from the origin, this would be an interesting non-SUSY, non-conformal field theory to study.)

At finite $k$, the CFTs are inherently strongly coupled, so one must resort to a bulk M-theory computation. The essential question of whether a nucleation instability \c{mms, switt} occurs boils down to whether the effective potential for probe M2-branes in AdS$_4\times S^7/\Z_k$, with the modified boundary condition for $\phi$ (dual to $\O$), is attractive or repulsive. (A related approach would be to compute the force between two probe M2-branes using the $\D=2$ boundary condition for $\phi$.) The only contributions to the potential that differ from the SUSY case must involve $\phi$ propagators. At leading order in $1/N$, one can compute the ``self-energy'' correction due to the emission and re-absorption of $\phi$ from the brane, given the $\D=2$ boundary condition. This should yield the leading order effect of the field theory potential \eqr{rg} in the IR. (A related calculation was performed in \c{craps}.) We expect that this leaves flat the $\la \O\ra=0$ subspace of the moduli space, so to determine whether this is lifted requires a higher-loop bulk computation. Unfortunately, this is no longer an AdS$_4$ supergravity computation, as it involves all of the scalar KK modes $\phi_p$, dual to $\O_p$. For instance, the emission/re-absorption of $\phi_p$ from the brane receives a loop correction, because all $\phi_p$ couple to $\phi$ through loops. %(e.g. the bulk coupling $\phi_p^2\phi$ is nonzero for all $p$, because $\phi$ sits in the supergravity multiplet.) 
%This loop induces mass shifts of $\phi_p$ due to the SUSY-breaking boundary condition for $\phi$. 
A vev for $\O_p$ would also seem to receive linear contributions from one-loop tadpole diagrams, where a $\phi$ loop attaches to a $\phi_p$ propagator attached to the probe; however, we note the encouraging feature that all cubic couplings $\phi_p\phi^2$ vanish.\foot{For $p>4$, this is just $\so(8)$ group theory (see \eqr{tensorp}); for $p=2,4$, while the coupling is allowed by group theory, it actually vanishes. We show this in Section \ref{4.1}.} There are likely other effects to consider as well; we leave a systematic exploration for future work.

\sec{Spectrum and Operator Products of the IR CFT}\label{s4}

The IR CFTs described here have a rigid structure despite the lack of SUSY. The spectrum of local operators is integer- or half-integer spaced (depending on the parity of $k$), to leading order in $1/N$. Their operator products obey strict selection rules imposed by $\so(8)$ global symmetry. The three-sphere free energy, $F=-\log Z_{S^3}$, 
can also be written to several subleading orders in $1/N$ \c{Klebanov:2011gs}:
\e{}{F_{IR} = F_{ABJM} - {\zeta(3)\o 8\pi^2} + \O(N^{-3/2})}
$F_{ABJM}$ is known exactly from the ABJM matrix model \c{marino}. In particular, it includes terms of order $N^{\pm 1/2}$ and $\log N$ that are identical in the IR CFT: higher loop effects start at $\O(N^{-3/2})$. It is somewhat remarkable that these subleading terms representing quantum effects in M-theory are robust to SUSY-breaking boundary conditions. 

Below we derive some new results on the CFT data at the putative IR fixed points. 

In Section \ref{4.1}, we show that even the {\it leading} large-$N$ contribution to certain single-trace correlators flows to zero in the IR, to leading order in $1/N$. In particular, this is true for ``extremal'' $n$-point correlators, and for the non-extremal three-point function $\la \O_2\O_2\O_2\ra$.\foot{In this Section, we revert to using $\O_2$ to label the superconformal primary in the ABJM stress tensor multiplet.}

In Section \ref{4.2}, we compute the change in the connected four-point functions $\la \O_p\O_p\O_p\O_p\ra$, with $p\neq 2$, between UV and IR. From this we extract, in Section \ref{4.3}, the leading-order change in the conformal dimensions of double-trace operators $\O_p \p^{2n}\p_{\mu_1}\ldots\p_{\mu_\ell}\O_p$ (modulo an issue of operator mixing, which we explain). However, the real power of this calculation comes in considering the consequences of $\mathfrak{so}(8)$ global symmetry. There exist several families of double-trace operators, one for each $\mathfrak{so}(8)$ representation appearing in the product $\O_p\otimes \O_p$. In the UV, some of these operators reside in SUSY-protected multiplets, and thus have vanishing anomalous dimensions to all orders in $1/N$. But in the IR, absent SUSY, these multiplets are no longer protected. Therefore, for these $\mathfrak{so}(8)$ representations, the {\it change} in the anomalous dimension between UV and IR {\it equals} the IR anomalous dimension! Thus, our computation allows us to read off some analytical double-trace spectral data about the IR CFT.

\ssec{Extremal correlators vanish after double-trace flow}\label{4.1}
An extremal correlator is defined by the condition
\e{hpext}{\Big\la \prod_{i=1}^n\O_{i}(x_i)\Big\ra~, \quad \text{where}\quad \D_1 = \sum_{i=2}^n \D_i}
These were studied mainly in the $\N=4$ SYM context in \c{extremal}, and then in the ABJM context in \c{0006103}. We now demonstrate two simple vanishing conditions of extremal correlators under double-trace flow, and of the non-extremal correlator $\la \O_2\O_2\O_2\ra$. 

\sssec{Three-point functions}
First, for completely general double-trace flows, an extremal three-point function that involves $\O$, the operator that triggers the flow, becomes zero in the IR to leading order in $1/N$:
\e{3pfe}{\text{\bf General double-trace flows:}\quad \la \O_i\O_j\O\ra_{UV} \neq 0 \quad \xrightarrow{\int \O^2}\quad  \la \O_i\O_j\O\ra_{IR} = 0}
for $\D_i=\D_j+\D_{\O}$. For the flow from ABJM, this implies that an infinite set of three-point functions vanishes at leading order in the IR: for all superconformal primaries $\O_p$ for any integer $p$ in the ABJM spectrum, 
\e{3ex}{\text{\bf ABJM:}\quad \la \O_{p+2}\O_p\O_2\ra = {8(p+1)\o \pi}\sqrt{\frac{2 (p+3)}{  (p+2)C_T}}\quad \xrightarrow{\int \O_2^2}\quad \la \O_{p+2}\O_p\O_2\ra_{IR} =0}
where the UV result may be read off from \c{bastianelli} (we have used a unit normalization of the two-point functions $\la \O_p\O_p\ra$).

It is straightforward to prove \eqr{3pfe}. For simplicity, consider the three-point extremal correlators $\la\P \O\O\ra$, where $\DP=2\D_\O$ in the UV. In the IR, we trade $\O$ for its Hubbard-Stratanovich field $\sigma$ \cite{gk} inside correlation functions. In Appendix \ref{appa}, we show that the IR OPE coefficient, $C_{\P\s\s}^{IR}$, is
\e{ext}{C^{IR}_{\P \s\s} = {C_{\P \O\O}^{UV}\o \pi^d C_{\O\O}^2}{\G({d\o 2}+{\DP\o 2}-\D_\O)\G^2(\D_\O)\o \G(\D_\O+{\DP\o2}-{d\o 2})\G^2({d\o2}-\D_\O)}{\G(d-{\DP\o2}-\D_\O)\o \G(\D_\O-{\DP\o2})}}
where $C_{\O\O}$ is the norm of $\O$. The denominator of the last factor implies that the UV-extremal correlator vanishes in the IR: $C^{IR}_{\P\s\s}=0$. The analogous calculation was done for $\la\P\Psi\O\ra$ where $\D_\P = \D_\Psi+\D_\O$ in \c{kirilin}, which, being extremal, can also be seen to vanish in the IR. 

This can be understood holographically using well-known facts about extremal correlators \c{extremal}. The bulk fields participating in extremal CFT three-point functions have a vanishing bulk cubic coupling, regardless of the boundary condition. In the UV, multiplying this zero by the infinity from the AdS integral gives a finite result (a more formal treatment involves subtle boundary terms). But in the IR -- that is, after changing quantization of the bulk field dual to $\O$ -- the AdS integral does not produce an infinity because the correlator is no longer extremal. Thus, one gets zero in the IR. This can also be understood yet another way, by thinking about operator mixing in the identification of bulk fields with CFT operators. A nonzero CFT extremal correlator $\la \P\O\O\ra$ is only consistent with a vanishing bulk coupling if the bulk field $\Phi_{\rm bulk}$ is dual not only to $\Phi$, but to a linear combination 
\e{}{\Phi_{\rm bulk} ~:=~ \P+{c\o \sqrt{C_T}} \O^2}
for some $c$ such that the three-point function of bulk modes, $\Phi_{\rm bulk}\O_{\rm bulk} \O_{\rm bulk}$, vanishes. In the IR, $\D_\O \rar d-\D_\O$, and this operator mixing is not allowed: thus, the holographic identification is $\Phi_{\rm bulk} :=\Phi$, and the CFT correlator vanishes.

\subsubsection*{Self-coupling of $\O_2$}
Let us also point out that the three-point function of $\O_2$ vanishes in the IR: again assuming unit normalization of $\O_2$,
\e{3ex}{\text{\bf ABJM:}\quad \la \O_2\O_2\O_2\ra = {128\o C_T}\quad \xrightarrow{\int \O_2^2}\quad \la \O_{2}\O_2\O_2\ra_{IR} =0}
This correlator is not extremal, but shares the feature that the bulk cubic vertex for $\phi_2$ vanishes; the nonzero result in the UV is due to a compensating factor $\G\left({\D_1+\D_2+\D_3-d \o 2}\right)$ in the AdS three-point scalar integrals \c{Freedman:1998tz}. (See \c{pw} for a proper treatment of boundary terms in $\N=8$ supergravity that yields the correct result.) In the IR where $\D\rar 2+\ldots$, this gamma function becomes finite, so the CFT three-point function vanishes. The analogous statement is true for the three-point function of $\phi^2$ in the large $N$ critical $O(N)$ model \cite{Petkou:1994ad,Petkou:2003zz, Sezgin:2003pt}. 

\sssec*{Application: Thermal one-point functions}
The fact that $\la \O_4\O_2\O_2\ra_{IR}=\la \O_2\O_2\O_2\ra_{IR}=0$ after flowing from ABJM modifies the leading large-$N$ behavior of thermal one-point functions in the IR. Consider first the one-point function of $\O_4$, defined on $S^1_{\b}\times S^2$ as
\e{}{ \la \O_4\ra_{S^1_\b\times S^2} = \text{Tr}_{\cal H}(\O_4e^{-\b H})}
where ${\cal H}$ is the local operator Hilbert space. The leading low-temperature asymptotics are determined by the dimension of the lightest operator to which $\O_4$ couples linearly. Thanks to the result above, the thermal one-point function of $\O_4$ has different behavior in UV and IR:
\es{}{{\rm UV:}\quad  \la \O_4\ra_{S^1_\b\times S^2}&\approx \la \O_2\O_4\O_2\ra e^{-\b}+\ldots\\
{\rm IR:}\quad  \la \O_4\ra_{S^1_\b\times S^2}&\approx \la \O_4\O_4\O_4\ra e^{-2\b}+\ldots}
The leading term in the IR comes from the cubic self-coupling of $\O_4$ because neither of the other IR operators with $\D\leq 2$ -- in particular, the $\D=3/2$ fermion and $\D=2$ scalar in the stress tensor multiplet -- produce a ${\bf 294_c}$ in their $\mathfrak{so}(8)$ tensor product \c{yamatsu}. Moreover, in the IR, all multi-trace operators made of $\O_2$ do not contribute to $ \la \O_4\ra_{S^1_\b\times S^2}$ at leading order in $1/N$: for these operators, the leading order contribution comes from (generalized) free field Wick contractions,
\e{}{\la [\underbrace{\O_2\ldots \O_2}_n]\,\O_4 \,[\underbrace{\O_2\ldots \O_2}_n]\ra_{IR} \approx \la \O_2\O_2\ra^{n-1}\la \O_2\O_4\O_2\ra_{IR}+\ldots~,}
which always leaves a three-point factor $\la \O_2\O_4\O_2\ra_{IR}=0$. 

Similar statements are true for the thermal one-point function of $\O_2$. For instance, at small $\b$,
\es{}{{\rm UV:}\quad  \la \O_2\ra_{S^1_\b\times S^2}&\approx \la \O_2\O_2\O_2\ra e^{-\b}+\ldots\\
{\rm IR:}\quad  \la \O_2\ra_{S^1_\b\times S^2}&\approx \la \psi\O_2\psi\ra e^{-{3\b\o 2}}+\ldots}
where $\psi$ is the spin-1/2 fermionic operator in the ${\bf 56}_v$ of $\so(8)$ (see Table \ref{tab1}). 

\sssec{$n$-point functions}

Next, in the double-trace flow from ABJM, any $n$-point extremal correlator involving at least one $\O_2$ also vanishes in the IR to leading non-trivial order in $1/N$:
\e{nex}{\text{\bf ABJM:}\quad \Big\la  \prod_{i=1}^{n-1}\O_{p_i}\O_2\Big\ra\neq 0 \quad \xrightarrow{\int \O_2^2}\quad  \Big\la  \prod_{i=1}^{n-1}\O_{p_i}\O_2\Big\ra_{IR} =0}
where $p_1 = 2+\sum_{i=2}^{n-1} p_i$. Although it is not directly to the question of whether tree-level extremal $n$-point correlators vanish after double-trace flow, we note for completeness that in ABJM (indeed, in maximally-SUSY CFTs in $3\leq d \leq 6$ \c{ferrara}), extremal correlators of chiral primaries exhibit the factorized form
%
%\e{hp2}{\Big\la \prod_{i=1}^n\O_{p_i}(x_i;Y_i)\Big\ra_{\rm ABJM} = \prod_{i=2}^n{(Y_1\cdot Y_i)^{p_i}\o |x_1-x_i|^{p_i}}}
\e{hp2}{\Big\la \prod_{i=1}^n\O_{p_i}\Big\ra_{ABJM} = \prod_{i=2}^n\la \O_{p_i}\O_{p_i}\ra}
where $p_1 = \sum_{i=2}^{n} p_i$ \c{0006103}. The mechanism can again be viewed as coming from the admixture of $\O_{p_1}$ with the $(n-1)$-trace operator $[\O_{p_2}\ldots \O_{p_n}]$. Upon flowing to the IR, this vanishes.

The proof of \eqr{nex} adapts the arguments of \c{extremal} to this setting. In fact, this was already done in \c{0006103}. For simplicity, we consider the $\N=8$ ABJM theories, so we study the four-point function 
\e{}{\la \O_{p_1}\O_{p_2}\O_{p_3}\O_{p_4}\ra~, \quad \text{where}\quad p_1=p_2+p_3+p_4}
For the double-trace application, we take (say) $p_4=2$. Using the $\so(8)$ tensor product (e.g. \c{ferrara})
\e{tensorp}{[00p_10] \otimes [00p_20] = \bigoplus_{k=0}^{p_2}\bigoplus_{j=0}^{p_2-k}[0,j,p_1+p_2-2k-2j,0]}
and likewise for $[00p_30] \otimes [00p_40]$, one sees that the only $\so(8)$ representation in both tensor products is $[00(p_3+p_4)0]$. The logic of \c{extremal} applies verbatim, and all bulk diagrams contributing to these correlators involve at least one vanishing bulk vertex. As explained earlier, it follows that the IR correlator vanishes.

\ssec{Four-point functions}\label{4.2}
In \c{kirilin}, the change in connected four-point functions $\la \P\P\P\P\ra$ was computed under general double-trace flows $\int \O^2$, for single-trace operators $\P$ that couple to $\O$. This was used to extract the change in the spectrum of double-trace operators $\P\p^{2n}\p_{\mu_1}\ldots \p_{\mu_\ell}\P$ in the IR, as well as their OPE coefficients. We can generalize this result to the present case, in which we take $\P$ to be $\O_p$, the superconformal primaries of ABJM.

\sssec{Setup}

The spectrum of superconformal primary operators of the ABJM theory, for any $k$, includes the infinite tower of 1/2-BPS chiral primaries $\O_p$, where $p=2,4,\ldots$, living in the $[00p0]$ representations of $\mathfrak{so}(8)$ or its branching into $\su(4)\times \u(1)$. These operators have conformal dimension $\D=p/2$. For concreteness, in the remainder of this section we specialize to $k=1,2$, and hence an $\so(8)$ global symmetry, though the results are easily generalized. 

We may form $\mathfrak{so}(8)$ invariants by contracting their indices with the null vectors $Y^I$,
\e{}{\O_p = \O_{I_1\ldots I_p} Y^{I_1}\cdots Y^{I_p}}
We will consider four-point functions $\la \O_p\O_p\O_p\O_p\ra$ for $p>2$, and compute their change under the RG flow triggered by \eqr{rg}: that is, under the deformation \eqr{rg}, we compute the quantity
\e{}{\la \O_p(x_1;Y_1)\O_p(x_2;Y_2)\O_p(x_3;Y_3)\O_p(x_4;Y_4)\ra_{IR} - \la \O_p(x_1;Y_1)\O_p(x_2;Y_2)\O_p(x_3;Y_3)\O_p(x_4;Y_4)\ra_{UV} }
to leading order. This is controlled, roughly speaking, by the order $g$ term in $\la \O_p\O_p\O_p\O_pe^{-g\int \O_2^2}\ra$. Due to SUSY Ward identities relating $\O_2$ to $T_{\mu\nu}$, $\O_2$ couples universally to all operators, so this difference is guaranteed to be nonzero for all $p$. The rest of the calculation is an extension of $\mathfrak{so}(8)$ group theory to the results of \c{kirilin}.  

The functional form of two- and three-point functions of $\O_p$ are determined by $\mathfrak{so}(3,2)\times \mathfrak{so}(8)$ symmetry. Let us introduce the following UV correlators,
\es{23pt}{\la \O_p(x_1;Y_1)\O_p(x_2;Y_2) \ra &= C_{pp}{(Y_1\cdot Y_2)^p\o x_{12}^p}\\
\la \O_p(x_1;Y_1)\O_p(x_2;Y_2) \O_2(x_3;Y_3)\ra &= C_{pp2}{(Y_1\cdot Y_2)^{p-1}(Y_2\cdot Y_3)(Y_3\cdot Y_1)\o x_{12}^{p-1}x_{23}x_{31}}}
for some constants $C_{pp}, C_{pp2}$, where $x_{12} \equiv |x_1-x_2|$. We may form the normalization-independent ratio,% which we compute in Appendix \ref{appb} as
\e{pp2}{a^{UV}_{pp2} = {C_{pp2}^2\o C_{pp}^2C_{22}}}
This may be computed in various ways (see e.g. \c{bastianelli}) to be
\e{}{a^{UV}_{pp2} = {32p^2\o C_T} }
Symmetry allows us to write the four-point function at either fixed point in the form
\e{521}{\la \O_p(x_1;Y_1)\O_p(x_2;Y_2)\O_p(x_3;Y_3)\O_p(x_4;Y_4)\ra = C_{pp}^2 \left({Y_1\cdot Y_2 Y_3\cdot Y_4\o x_{12}x_{34}}\right)^p \F_p(u,v;\sigma,\tau)}
where we introduced the internal cross-ratios
\e{}{\sigma = {Y_1\cdot Y_3Y_2\cdot Y_4\o Y_1\cdot Y_2Y_3\cdot Y_4}~, ~~ \tau = {Y_1\cdot Y_4Y_2\cdot Y_3\o Y_1\cdot Y_2Y_3\cdot Y_4}}
$\F_p(u,v;\sigma,\tau)$ has an expansion in the $\mathfrak{so}(8)$ representations appearing in the tensor product $\Rc_p\otimes \Rc_p$, where we sometimes employ the shorthand
\e{}{\Rc_p \equiv [00p0]}
The list of such representations is 
\es{ab}{\Rc_p\otimes\Rc_p &= \bigoplus_{a=0}^p\bigoplus_{b=0}^a\, [0(a-b)(2b)0] \\
&\equiv \bigoplus_{a=0}^p\bigoplus_{b=0}^a (ab)}
Representations in the symmetric product have $a+b =$ even, whereas those in the anti-symmetric product have $a+b=$ odd. The contribution of each representation to $\d\F_p$ is encoded in a harmonic polynomial of $\mathfrak{so}(8)$ which depends on both $\sigma$ and $\tau$ \c{osborn, dolan}. These polynomials $Y_{ab}(\s,\t)$, associated to the representation $(ab)$, obey an orthogonality condition
\e{ortho}{\int\!\!\!\int Y_{ab}(\s,\tau) Y_{cd}(\s,\tau) \propto \delta_{ac}\delta_{bd}}
with 
\e{}{\dint \equiv \int_0^{(1-\sqrt{\tau})^2} d\sigma \int_0^1 d\tau\, \left[ (\s-1)^2+\tau(\tau-2\sigma-2)\right]^{3/2}}
Hence the four-point function enjoys the decomposition
\e{526}{\F_p(u,v;\sigma,\tau) = \sum_{a,b} Y_{ab}(\sigma,\tau) f_{ab}(u,v)~, ~~\text{where}~~ \sum_{a,b}\equiv \sum_{a=0}^p\sum_{b=0}^a}
Crossing symmetry of $\F_p(u,v;\sigma,\tau)$ acts in all four variables. The algorithm for constructing these polynomials, as well as the first several explicit polynomials, can be found in \cite{osborn, dolan}. Of particular importance for what follows are the polynomials associated to $(00)=[0000]$, the singlet, and $(11)=[0020]$, in which $\O_2$ lives:
\es{y0011}{Y_{00}(\s,\t)&=1\\
Y_{11}(\s,\t) &= \s+\t-{1\o4}}

In what follows, we will compute $\d\F_p(u,v;\s,\t)$, the difference between IR and UV connected correlators:
\e{}{\d\F_p(u,v;\s,\t) \equiv \F_p(u,v;\s,\t)\big|_{\rm IR} - \F_p(u,v;\s,\t)\big|_{\rm UV}}
\sssec{Change in four-point function}
As explained in \c{kirilin}, $\d\F_p$ is given by a sum of three terms, each of which computes the change in the contribution of $\O_2$ in a given channel. Each contribution carries the $R$-symmetry polynomial $Y_{11}(\s,\t)$ in its respective channel. This is the only $\mathfrak{so}(8)$ representation that appears in a given channel, because we are taking the difference of the four-point functions at the two fixed points. Combining this global symmetry structure with the explicit result of \c{kirilin}, we find 
\es{abjmres}{\d \F_p(u,v;\sigma,\tau) = -{16p^2\o \pi^{5/2}C_T}&\Big[Y_{11}(\s,\tau) u\bar{D}_{1,1,{1\o 2}, {1\o 2}}(u,v)\\&+\s^p \,Y_{11}(\s^{-1},\tau\s^{-1})u^{p\o 2} \bar{D}_{1, {1\o 2}, 1, {1\o 2}}(u,v) \\&+\tau^p \,Y_{11}({\s\tau^{-1}},\tau^{-1})\left(\frac{u}{v}\right)^{p\o 2} v \bar{D}_{{1\o 2}, 1, 1, {1\o 2}}(u,v)\Big]}
where the $\bar D$-function is defined by the integral
\es{}{&\int d^d y\prod_{i=1}^4\frac{\G(\D_i)}{(x_{i}-y)^{2\Delta_i}}\stackrel{\sum \Delta_i = d}{=}\pi^{\frac{d}{2}} \,
\frac{x_{14}^{d-2\Delta_1-2\Delta_4} x_{34}^{d-2\Delta_3-2\Delta_4}}{x_{13}^{d-2\Delta_4} x_{24}^{2\Delta_2}}
\bar{D}_{\Delta_1\Delta_2\Delta_3\Delta_4}(u,v)}
and we have used the OPE coefficients \eqr{pp2}. Each line of \eqr{abjmres} represents a different OPE channel. In writing this, we have used the obvious transformation properties of $\sigma$ and $\tau$ under permutation of the indices, together with \eqr{521}.\foot{The first line of \eqr{abjmres} is $Y_{11}(\s,\t)/2$ times the result one would obtain for the same double-trace flow without global symmetries (likewise for the other two channels). The factor of 1/2 comes from contracting the vectors $Y_{1,2,3,4}$ with the tensor structure $\d_{IK}\d_{JL}+\d_{IL}\d_{JK}-\d_{IJ}\d_{KL}/4$ that appears in the two-point function of the Hubbard-Stratanovich field for $\O_2$, and using the normalization \eqr{y0011}. See \c{kirilin} for details.}

Eq. \eqr{abjmres} is the complete result. It is useful to project $\d\F_p$ into a single OPE channel -- say, the $12\rar34$ channel -- by putting it in the form \eqr{526}. This makes it straightforward to extract anomalous dimensions for the double-trace operators $[\O_p\O_p]$.  To do so, we project \eqr{abjmres} onto each representation $(ab)$. Let us define a normalized projection operator $\Pc_{ab|cd}(p)$, that projects $Y_{cd}$ polynomials in the $t$-channel onto $Y_{ab}$ polynomials in the $s$-channel:
\e{projdef}{\Pc_{ab|cd}(p) \equiv {1\o\N_{ab}} \dint \s^p\,Y_{ab}(\s,\tau) \,Y_{cd}(\s^{-1},\tau\s^{-1})}
where $\N_{ab}$ is the norm,
\e{}{\N_{ab} = \dint Y_{ab}(\s,\tau)^2}
The $u$-channel projection is identical up to a $(-1)^{a+b}$. Applied to \eqr{abjmres}, this projector acts on the $(\s,\tau)$-dependent parts of the second two lines, with $(cd)=(11)$: one finds
\es{chan}{\d\F_p&(u,v;\sigma,\tau) =  -{16p^2\o \pi^{5/2}C_T}\Bigg\lbrace Y_{11}(\s,\tau) u\bar{D}_{1,1,{1\o 2}, {1\o 2}}(u,v)\\&+\sum_{a,b} Y_{ab}(\s,\tau)\Pc_{ab|11}(p)\left[u^{p\o 2} \bar{D}_{1, {1\o 2}, 1, {1\o 2}}(u,v)+(-1)^{a+b}\left(\frac{u}{v}\right)^{p\o 2} v \bar{D}_{{1\o 2}, 1, 1, {1\o 2}}(u,v)\right]\Bigg\rbrace}
The factor of $(-1)^{a+b}$ indicates whether the $(ab)$ representation appears in the symmetric $(+)$ or anti-symmetric $(-)$ product $\Rc_p\otimes\Rc_p$.

\ssec{Double-trace anomalous dimensions}\label{4.3}
The first line of \eqr{chan} represents the exchange of $\O_2$, while the second line of \eqr{chan}  represents the exchange of double-trace operators of the schematic form
\e{}{[\O_p\O_p]_{n,\ell}^{(ab)} \simeq \O_p \p^{2n}\p_{\mu_1}\ldots\p_{\mu_\ell}\O_p\Big|_{(ab)\, \subset\, \Rc_p\otimes \Rc_p}}
The notation indicates that there exist several families of such operators, one for each $\mathfrak{so}(8)$ representation appearing in the product $\Rc_p\otimes \Rc_p$. We denote their total conformal dimension as $\D_{n,\ell}^{(ab)}(p)$, and introduce an anomalous dimension
\e{anom}{\g_{n,\ell}^{(ab)}(p) \equiv \D_{n,\ell}^{(ab)}(p)-(2\D_p+2n+\ell)}
The $\g_{n,\ell}^{(ab)}(p)$ have a $1/C_T$ expansion; we will henceforth take $\g_{n,\ell}^{(ab)}(p)$ to be the leading term, of order $1/C_T$, dual to tree-level contributions to the binding energy in AdS. We focus on the leading-twist operators, with $n=0$, and introduce the shorthand $\g_\ell\equiv \g_{0,\ell}$.

By decomposing the second line of \eqr{chan} into double-trace conformal blocks and working in the $1/C_T$ expansion, we can extract the flow of anomalous dimensions from UV to IR. Define
\e{dguvir}{\d\g_{\ell}^{(ab)}(p) \equiv \g_{\ell}^{(ab)}(p)\big|_{\rm IR} - \g_{\ell}^{(ab)}(p)\big|_{\rm UV}~.}
In Appendix \ref{appb}, we carry out the remaining steps in the calculation. The result for even $p$ is
\e{ad1}{\boxed{\d\g_{\ell}^{(ab)}(p) = \d\g_0^{(ab)}(p) \sum_{k=0}^{{p-4\o 2}}{c_k(p)\o \ell+{p\o 2}+k}}}%~~\text{where}~~\ell\in 2\Z_{\geq 0}}
where
\e{ckform}{c_k(p) = \frac{(p-2) \left(2-\frac{p}{2}\right)_{k} \left(\frac{p-1}{2}\right)_{k}}{(p-3) \left(\frac{5}{2}-\frac{p}{2}\right)_{k} \left(\frac{p}{2}\right)_{k}}}
and
\e{ad2}{\d\g_{0}^{(ab)}(p) =  \frac{64p^2}{\pi^2 C_T}{\Pc_{ab|11}(p)\o \Pc_{ab|00}(p)}}
where $\ell$ is even/odd if $a+b$ is even/odd. (One can easily check that the sum on the RHS of \eqr{ad1} equals unity for $\ell=0$.) In Appendix \ref{appb}, we also give the result for odd $p$, and the explicit functions $\Pc_{ab|00}(p)$ and $\Pc_{ab|11}(p)$ for $a,b\leq 4$ and $a+b=$ even. Note that the ratio of $\d\g_\ell^{(ab)}$ for two different spins is completely independent of the $\mathfrak{so}(8)$ representation $(ab)$. 
\vs
Actually, we should note that this result for $\d\g_\ell^{(ab)}(p)$ does not take into account large $N$ operator mixing in ABJM. At leading order in $1/N$, the following set of double-trace operators in the ABJM theory have the same conformal dimensions and spins, and therefore undergo mixing:
\e{mixing}{[\O_p\O_p]^{(ab)}_{n,\ell}~, ~~ [\O_{p-2}\O_{p-2}]^{(ab)}_{n+1,\ell}~, ~~ [\O_{p-4}\O_{p-4}]^{(ab)}_{n+2,\ell}~, \ldots~,}
All of these operators must sit in the same $\so(8)$ representation $(ab)$. Therefore, \eqr{ad1} should be viewed as a linear combination of contributions from all such operators to the intermediate channel of $\la \O_p\O_p\O_p\O_p\ra$, weighted by their squared OPE coefficients. The full mixing problem is complicated, and we do not solve it here.\foot{See the recent work \c{drummond}, and \c{aldaybissi}, for the solution of the analogous mixing problem in $\N=4$ SYM. We thank J. Drummond for reminding us of the mixing problem.} For application in the next subsection, note the simplifying feature that, due to the structure of $\so(8)$ tensor products in \eqr{ab}, there is no mixing when $a=p-1,p$: only the $[\O_p\O_p]^{(ab)}_{n,\ell}$ operator can sit in these representations. 

Let us make a comment on signs. By inspection, $c_k(p)>0$ for all $p\geq 4$, so the sign of $\d\g_\ell^{(ab)}(p)$ is given by the sign of the ratio of projectors. In general, these ratios need not be sign-definite: whereas $\Pc_{ab|00}(p)>0$ due to unitarity of mean field theory (see \eqr{b9}), there is no unitarity constraint on $\Pc_{ab|cd}(p)$ for $(cd)\neq (00)$. Explicit calculation using the projectors in \eqr{proj00}-\eqr{proj11} does in fact produce both signs for different representations at fixed $p$.\foot{Both signs are consistent with lightcone bootstrap constraints on large spin anomalous dimensions, due to the non-trivial global symmetry representations involved. See e.g Section 2 of \c{poland} for similar examples of charged correlators, there studied in the $\ell\gg1$ limit, where an intricate pattern of signs was found.} For instance, in the case $p=4$, for representations appearing in the symmetric product $[\Rc_4\otimes \Rc_4]_{\rm sym}$, one finds
\es{}{\d\g_{0}^{(ab)}(p=4)&>0 ~~\forall~~ (ab)\in \lbrace (00), (11), (20), (22), (31), (33), (44)\rbrace \\
\d\g_{0}^{(ab)}(p=4)&<0 ~~\forall~~ (ab)\in \lbrace (40), (42)\rbrace }
This pattern appears to generalize to $p\neq 4$: the only symmetric representations for which $\d\g_\ell^{(ab)}<0$ are those with $a=p, b<p$. It would be nice to prove this.

%\vs
\sssec{IR dimensions for UV-protected operators}\label{s432}
In the tensor product $\Rc_p\otimes \Rc_p$, given in \eqr{ab}, the operators living in representations $(ab)$ with $a=p-1,p$ are protected by SUSY in the UV; all others are unprotected \c{dolan}. For this subset of protected representations, our double-trace data is especially interesting: since $\g_{n,\ell}^{(ab)}(p)\big|_{\rm UV}=0$, the change in anomalous dimension under RG flow equals the IR anomalous dimension. 

There are further constraints from $\mathfrak{osp}(8|4)$ representation theory on which composites are protected. In Table \ref{tab2} we show the relation between internal and spacetime quantum numbers for the protected $\so(8)$ multiplets in the $(ab)$ representations. Because $\D_{n,\ell}^{(ab)}(p) = p+2n+\ell$ for protected representations, Table \ref{tab2} implies the following: 

\begin{table}[]
\begin{center}
 \begin{tabular}{|c|c|c|}
\hline
Class& $\tau_{\O}$ & $\ell_\O$   \\
  \hline
$ A$ &$a+1$& $\geq 0$   \\
$ B $&$a$ & $0$   \\
\hline
 \end{tabular}
\end{center}
\caption{The two classes of BPS multiplets of $\so(8)$, specified to the $(ab)\equiv [0(a-b)(2b)0]$ representations, along with the twist $\tau_\O = \D_\O-\ell_\O$ and Lorentz spin $\ell_\O$ of the superconformal primary $\O$.  See e.g. \c{cordova} or Table 2 of \c{pufu} for further refinement.}
\label{tab2}
\end{table}%

\begin{itemize}

\item When $a=p$, the protected superconformal primary double-trace operators lie in the $B$ series of BPS representations, with $\ell=0$.  

\item When $a=p-1$, the protected superconformal primary double-trace operators lie in the $A$ series of BPS representations, with $b-\ell=$ even.

\item Among superconformal primaries, only the $n=0$ operators are protected.\foot{These operators have conformal primary descendants, which are also UV-protected; these can be easily enumerated by expanding the supermultiplet operator content. Such conformal primaries may have $0\leq n \leq 3$, depending on how many supercharges generate the full multiplet. For $n>0$, there is mixing among double-trace operators of $n'\leq n$ and identical spins.}
\end{itemize}

\noindent As explained below \eqr{mixing}, there is no operator mixing for these representations.
 Thus, equations \eqr{ad1}-\eqr{ad2} give analytic formulas for the leading order anomalous dimensions of infinite classes of double-trace operators at the non-SUSY IR fixed point: for the values of $(n,\ell)$ noted above, 
\e{prot}{\boxed{~\d\g_{n,\ell}^{(ab)}(p) = \g_{n,\ell}^{(ab)}(p)\Big|_{\rm IR} ~~\text{when}~~ a=p-1~\text{or}~ p\,.~}}

\vs

For example, let us provide the explicit IR dimensions of the infinite class of scalar double trace primaries in the symmetric traceless representation $(pp)=[00(2p)0]$, with $n=0$:
\e{pp}{[\O_p\O_p]_{0,0}^{(pp)}  \equiv \,\,:\!\O_p\O_p\!:\!\Big|_{(pp)}} 
In the UV, these operators are superconformal primaries living in a 1/2-BPS $B$ series multiplet, with vanishing anomalous dimension. In the IR, the anomalous dimension %ir conformal dimensions are
$\g_0^{(pp)}(p)$, as defined in \eqr{anom}, is given in \eqr{ad2}. By inspection of the projectors through $p=18$, we find that for this representation, the ratio of projectors appearing in \eqr{ad2} is actually $p$-independent:
\e{34p}{{\Pc_{pp|11}(p)\o \Pc_{pp|00}(p)} = {3\o 4}~.}
This leads to a particularly simple result for the leading-order IR anomalous dimension of \eqr{pp},
\e{}{\g_0^{(pp)}(p)\Big|_{\rm IR} = {48p^2\o\pi^2 C_T}}

As another example, let us also, using \eqr{proj00}-\eqr{proj11}, give the explicit IR anomalous dimensions of the leading-twist $p=4$ spinning double-trace operators,
\e{p4}{[\O_4\O_4]_{0,\ell}^{(ab)}  \equiv \,\,:\!\O_4\p_{\mu_1}\ldots \p_{\mu_\ell}\O_4\!:\!\Big|_{(ab)}~,} 
in the symmetric $(3b)$ and $(4b)$ representations:
\begin{equation}
\setlength{\jot}{6pt} % affecting the line spacing in the environment
\begin{split}
\g_\ell^{(31)}(4)\Big|_{\rm IR}~ &= ~{3328\o3\pi^2 C_T}{1\o \ell+2}\\
\g_\ell^{(33)}(4)\Big|_{\rm IR} ~ &= ~ {1536\o\pi^2 C_T}{1\o \ell+2}\\
\g_0^{(40)}(4)\Big|_{\rm IR} ~ &= ~ -{512\o\pi^2 C_T}\\
\g_0^{(42)}(4)\Big|_{\rm IR} ~ &= ~ -{128\o\pi^2 C_T}\\
\g_0^{(44)}(4)\Big|_{\rm IR} ~ &= ~ {768\o\pi^2 C_T}
\end{split}
\end{equation}
where $\ell\in2\Z_{\geq 0}$.  

These results for $\g_\ell^{(ab)}(p)$ at the IR fixed point are the first analytic computations of anomalous dimensions of finite-spin double-trace operators in any large $N$ CFT$_3$ with an Einstein gravity dual. The only previously known data at finite spin, either analytic or numeric, is a numerical bootstrap estimate in $\N=8$ ABJM for the $\so(8)$-singlet operators $[\O_2\O_2]_{0,\ell}$ for $\ell=0,2$ \c{pufu}. (Large spin results have been obtained using the lightcone bootstrap \c{fitz, kz, az, poland2, poland}.) This technique could also be applied to derive anomalous dimensions in IR CFTs obtained by double-trace RG flows that {\it preserve} a fraction of the UV SUSY: again, certain UV-protected double-trace operators become unprotected in the IR, as determined by the branching rules of the UV superalgebra. %The above results are also, to our knowledge, the first calculations of $\g_\ell$ at finite spin for double-trace operators in non-trivial global symmetry representations, in any large $N$, large gap CFT in any $d$. 
 The interpretation of CFT anomalous dimensions as AdS binding energies has been discussed elsewhere \c{Cornalba:2006xm, Fitzpatrick:2010zm, Fitzpatrick:2011hh}.

\sec{Final comments}\label{s5}
The proposal we have made, and the specific example involving ABJM, provide a way to construct non-SUSY CFTs with large $N$ and a large gap that appear to obey all necessary CFT consistency conditions. Of course, it is paramount to understand if there is nevertheless an obstruction. It would be enlightening, though challenging, to fully determine the fate of the moduli space of vacua after the RG flow. If instabilities do develop, the conjecture of \c{OV, FK} will have passed a novel test; if they do not, a plausible modification of the conjecture is that {\it all} non-SUSY CFTs with a large gap are obtained by SUSY-breaking RG flows. This is still a radical statement that, if true, would be fascinating from the CFT perspective: in the absence of SUSY, the typical large $N$, large gap CFT is believed to be complicated, with a highly disordered set of irrational operator dimensions and OPE coefficients and a host of possible sporadic phenomena. On the other hand, CFTs constructed via double-trace flow are highly ordered. 

We have only computed a handful of gauge-invariant observables of the putative IR fixed point obtained by flowing from ABJM, but it is worth exploring its properties further. For instance, one would like to compute the leading-order shift of the single-trace spectrum in the IR, where $\O_p$ may acquire anomalous dimensions. We may do so by an AdS computation of the one-loop correction to the propagators of KK modes $\phi_p$ on AdS$_4\times S^7/\Z_k$. The relevant bubble and tadpole diagrams -- see Figure \ref{fig4} -- 
can be computed as explained e.g. in \c{Giombi:2017hpr}. More precisely, to compute the IR dimensions of $\O_p$, 
it would be sufficient to compute differences of such diagrams with $\Delta=1$ and 
$\Delta=2$ boundary conditions on $\phi_2$. A missing ingredient are the quartic couplings $\phi_2^2\phi_p^2$, which have never been computed. 
These would also allow computation of the four-point functions in the large $N$ ABJM theory itself, which remains an outstanding 
problem.\footnote{In fact, note that the calculation of differences of loop diagrams with $\Delta=2$ and $\Delta=1$ boundary conditions on $\phi_2$ 
can be mapped, following \cite{HartmanRastelli, GiombiYin}, 
to a conformal perturbation theory computation on the CFT side. This includes
a contribution proportional to $\int dz_1 dz_2 G_{\sigma}(z_1,z_2) \langle \O_p(x_1) \O_p(x_2) \O(z_1) \O(z_2)\rangle_{UV} $, 
which entails knowing the four-point function in the ABJM theory in the UV.} 
 \begin{figure}
    \centering
       \includegraphics[width = .6\textwidth]{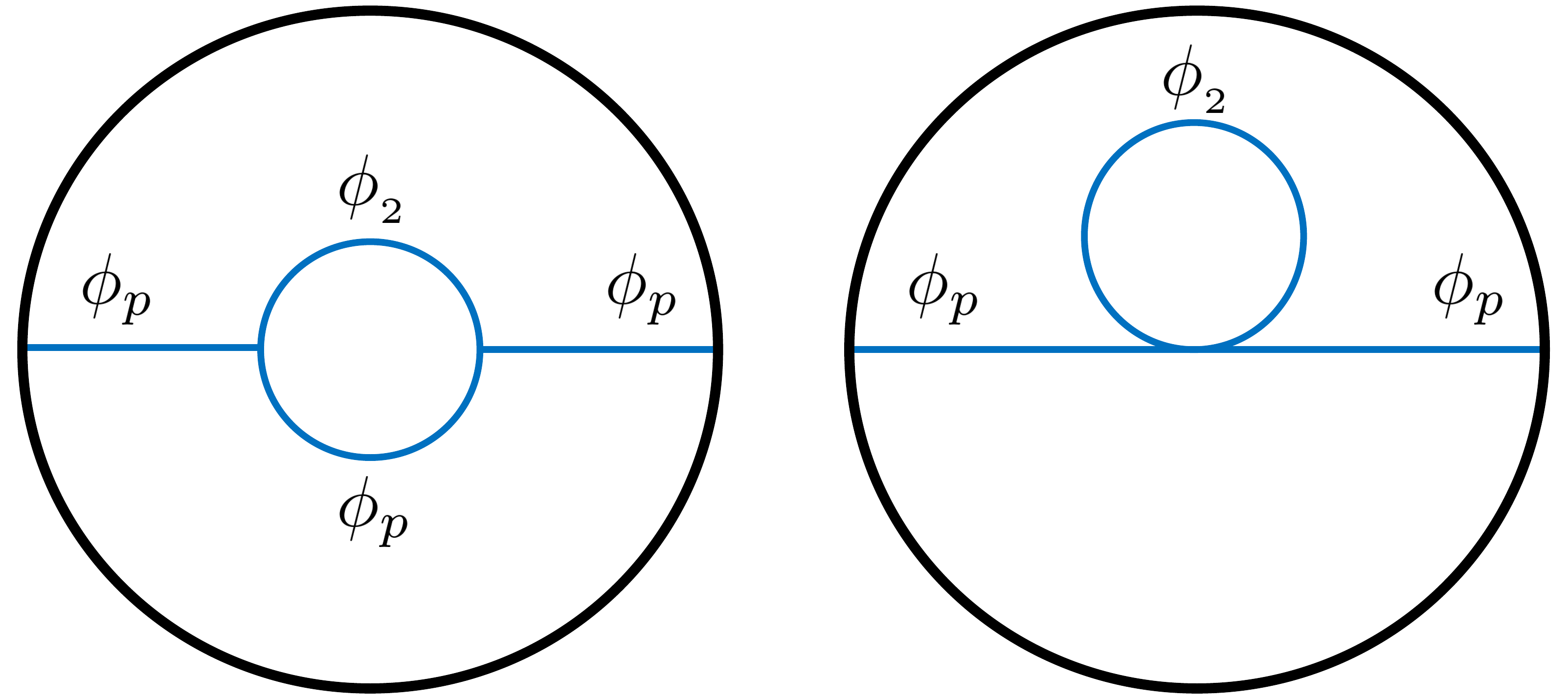}
         \caption{The bulk diagrams needed to determine the leading anomalous dimensions of the single-trace operators $\O_p$ after the ABJM double-trace flow by $\int\! \O_2^2$.}
         \label{fig4}
\end{figure}

An intriguing question, independent of the concerns of this paper, is whether the RG flow \eqr{rg} survives all the way down to small values of $N$. Can one reach an analog of the Wilson-Fisher model, endowed with $\so(8)$ global symmetry, by RG flow from ABJM? Away from large $N$, the notion of ``double-trace flow'' is meaningless, but \eqr{rg} can be understood as a fancier version of the typical $\phi^4$ deformation, in analogy with the usual construction of the Wilson-Fisher fixed point via RG flow from the free scalar theory. Such a CFT, if it exists, may (but need not) be a non-SUSY Chern-Simons-bifundamental matter theory. Recent studies of non-SUSY Chern-Simons-matter theories have revealed a rich landscape of fixed points and dualities (e.g. \c{Giombi:2011kc, Aharony:2011jz, Karch:2016sxi, Murugan:2016zal, Seiberg:2016gmd}); it would be interesting to ask whether this landscape accommodates the $\so(8)$ Wilson-Fisher-type theory described above. One promising approach to this problem may be to ask the conformal bootstrap whether such a theory is allowed to exist, for instance, by generalizing the analysis of \c{pufu} to include $\so(8)$ global symmetry but not SUSY.\foot{A preliminary problem is to understand how many relevant singlet operators there are in the ABJM theories at some finite $N$ and $k$. At large but finite $N$, the answer is at most two: the finite $N$ continuations of the double- and triple-trace operators $[\O_2\O_2]$ and $[\O_2\O_2\O_2]$ projected onto the $R$-symmetry singlet. (The leading-order anomalous dimension of this triple-trace operator has not been computed; in particular, its sign is not known.) As we decrease $N$ further, singlet single-trace operators such as ${\rm Tr}(X_IX_I)$, the ABJM analog of the Konishi operator, re-enter the spectrum. For $k=1,2$, it should be possible to extend the $\N=8$ numerical bootstrap methods of \c{pufu} to determine the number of relevant operators.}

One might also try to construct SUSY-breaking double-trace flows from large-gap SCFTs in $d=2$. It behooves us to look for more M-theory examples. A canonical one is M-theory on AdS$_3\times S^2\times CY_3$, whose dual is the MSW CFT with $\N=(0,4)$ SUSY \c{msw}. This theory remains poorly understood, but the BPS spectrum is known \c{Larsen:1998xm, Fujii:1998tc, deBoer:2009un}, and contains no $\D<1$ operators. It would be worthwhile to seek other examples, particularly given the paucity of explicit constructions of large $N$ CFTs in $d=2$ with sparse spectra. 

\section*{Acknowledgments}

We thank O. Aharony, A. Armoni, S. Chester, J. Drummond, D. Jafferis, I. Klebanov, P. Kraus, H. Ooguri, S. Pufu, L. Rastelli and H. Verlinde for helpful discussions, and V. Kirilin for collaboration on related work. We also thank I. Klebanov and H. Ooguri for comments on a draft. The work of S.G. is supported in part by the US NSF under Grant No.~PHY-1620542. E.P. gratefully acknowledges support from the Simons Center for Geometry and Physics, Stony Brook University at which some of the research for this paper was performed. E.P. is supported in part by the Department of Energy under Grant No. DE-FG02-91ER40671, and by Simons Foundation grant 488657 (Simons Collaboration on the Nonperturbative Bootstrap). This material is based upon work supported by the U.S. Department of Energy, Office of Science, Office of High Energy Physics, under Award Number DE-SC0011632.

\appendix

\begin{figure}
    \centering
       \includegraphics[width = .4\textwidth]{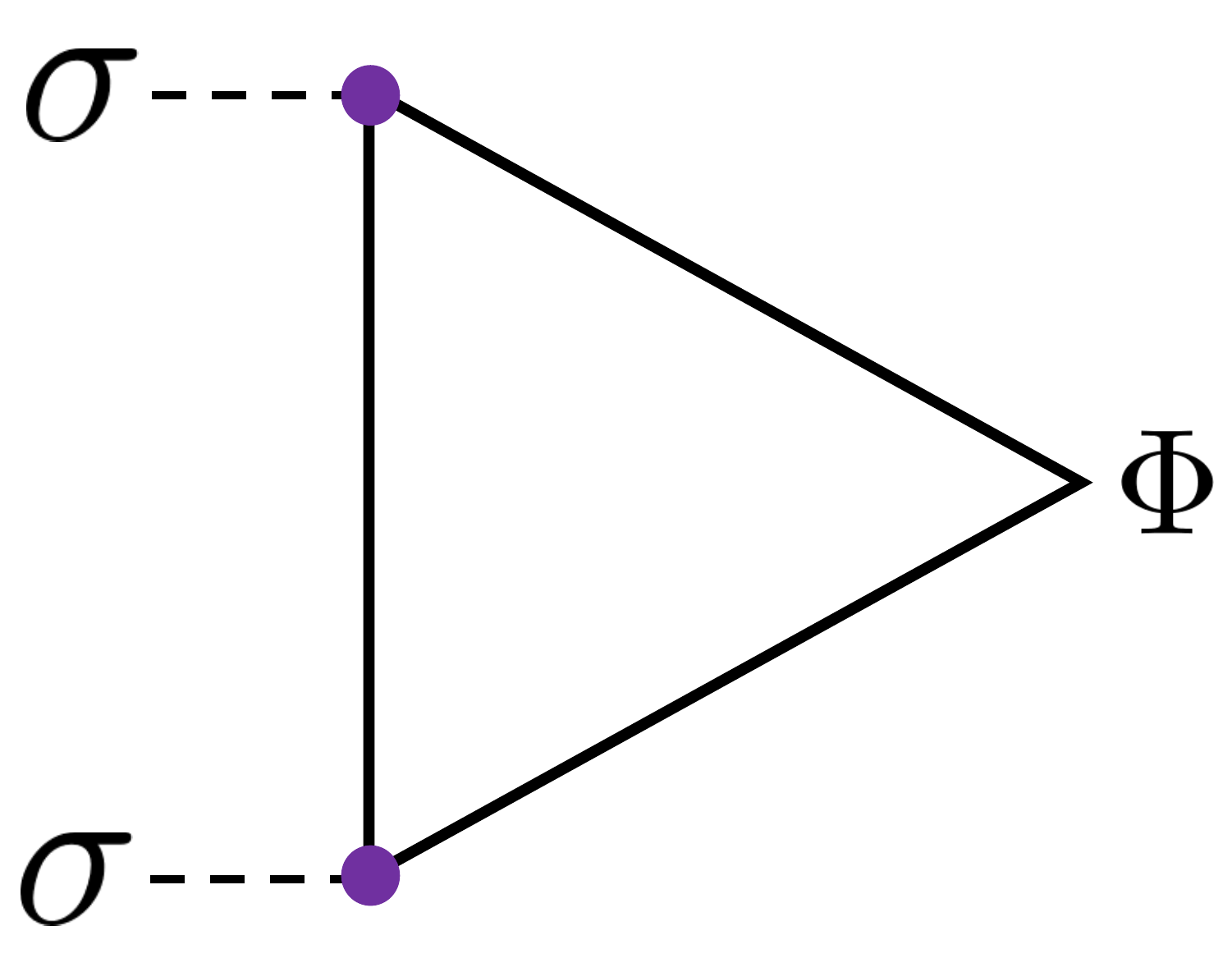}
         \caption{The triangle diagram determines the three-point coupling $\la \P\sigma\sigma\ra$ in the IR, to which the UV coupling $\la \P\O \O\ra$ flows. The purple points are integrated over.}\label{fig1}
\end{figure}

\sec{Extremal three-point functions under double-trace flow}\label{appa}
Consider the extremal three-point function 
\e{}{\la \P(x_1)\O(x_2)\O(x_3)\ra = {C_{\P \O\O}\o x_{12}^{\DP}x_{13}^{\DP}}~, ~~\text{where}~~ \DP=2\D_\O}
We now perturb the CFT by $\int\! \O^2$. To leading order in $1/N$, in the IR we take $\O\rar\s$ inside correlation functions, whereupon we must compute the ``triangle diagram'' with two $\s$ legs; see Figure \ref{fig1}. $\s$ has a two-point function \cite{gk, Diab:2016spb}
\e{}{\la \s(x_1)\s(x_2)\ra = \frac{({d\o2}-\D_\O)\sin\left(({d\o2}-\D_\O)\pi\right)\Gamma\left(d-\D_\O\right)\Gamma\left(\D_\O\right)}
{\pi^{d+1}C_{\O\O}x_{12}^{2(d-\D_\O)}}\equiv \frac{C_{\sigma}}{x_{12}^{2(d-\D_\O)}}}
where $x_{ij} \equiv |x_i-x_j|$, and 
\e{}{\la \O(x_1)\O(x_2)\ra = {C_{\O\O}\o x_{12}^{2\D_\O}}}
The necessary integral is
\es{}{\la \P(x_1)\s(x_2)\s(x_3)\ra_{IR} &= \int d^d x_4 \int d^d x_5 {C_\s\o x_{24}^{2(d-\D_\O)}}{C_\s\o x_{35}^{2(d-\D_\O)}}\la \P(x_1)\O(x_4)\O(x_5)\ra_{UV}+\ldots\\
&= \int d^d x_4 \int d^d x_5 {C_\s\o x_{24}^{2(d-\D_\O)}}{C_\s\o x_{35}^{2(d-\D_\O)}} {C_{\P\O\O}^{UV}\o x_{14}^{\DP}x_{15}^{\DP}}+\ldots}
where $\ldots$ denotes higher orders in $1/N$, and we have used conformal symmetry to go from the first to the second line. Two applications of the conformal integral
\e{}{\int \frac{d^d x_4}{x_{14}^{2\Delta_1}x_{24}^{2\Delta_2}x_{34}^{2\Delta_3}}~\stackrel{\sum \Delta_i = d}{=}~\frac{\pi^{\frac{d}{2}}a(\Delta_1)a(\Delta_2)a(\Delta_3)}{x_{12}^{d-2\Delta_3}x_{23}^{d-2\Delta_1}x_{31}^{d-2\Delta_2} }\,,}
where
\e{}{a(\Delta_i)\equiv\frac{\Gamma(d/2-\Delta_i)}{\Gamma(\Delta_i)}~,}
lead to the result \eqr{ext} quoted in the text. One concludes that UV-extremal correlators involving $\O$ vanish in the IR. Note that the reverse is also true: if the correlator is not extremal in the UV, but becomes extremal in the IR -- that is, if $\DP=2(d-\D_\O)$ -- the numerator of the last factor in \eqr{ext} blows up; this gives a finite result only if $C_{\P\O\O}^{UV}=0$.

\sec{Double-trace computations and $\mathfrak{so}(8)$ group theory}\label{appb}

\ssec{$\mathfrak{so}(8)$ projectors}
In \eqr{projdef}, we introduced the projector $\Pc_{ab|cd}(p)$, whose definition we recall here:
\e{}{\Pc_{ab|cd}(p) \equiv {1\o \N_{ab}}\dint \s^p \,Y_{ab}(\s,\tau)Y_{cd}(\s^{-1},\t\s^{-1})}
where
\e{}{\dint \equiv \int_0^{(1-\sqrt{\tau})^2} d\sigma \int_0^1 d\tau\, \left[ (\s-1)^2+\tau(\tau-2\sigma-2)\right]^{3/2}}
and
\e{}{\N_{ab} \equiv \dint Y_{ab}(\s,\tau)^2}
This projects a $t$-channel exchange in the $(cd)$ representation of $\mathfrak{so}(8)$ of the correlator $\la \O_p\O_p\O_p\O_p\ra$, where $\D_p = p/2$, onto the $s$-channel. The identical formula, up to an overall $(-1)^{a+b}$, holds for projection of a $u$-channel term onto the $s$-channel. A nice exposition of the polynomials $Y_{ab}(\s,\t)$, and a list of those with $a\leq 3$, is given in Appendix B of \c{osborn}. For what follows, we will also need
\es{}{Y_{40}(\s,\t) &= \left(\s^4-4 \s^3 \t+6 \s^2 \t^2-4 \s \t^3+\t^4\right) -\frac{6}{5}  \left(\s^3-\s^2 \t-\s \t^2+\t^3\right)\\&+\frac{3}{110} \left(17 \s^2-12 \s \t+17 \t^2\right) -\frac{3}{55}  (\s+\t)+\frac{1}{330}\\
Y_{42}(\s,\t) &=  \left(\s^4+2 \s^3 \t-6 \s^2 \t^2+2 \s \t^3+\t^4\right)+\frac{1}{4} \left(-5 \s^3-3 \s^2 \t-3 \s \t^2-5 \t^3\right)\\&+\frac{3}{44} \left(7 \s^2+6 \s \t+7 \t^2\right)+-\frac{3}{44} (\s+\t)+\frac{1}{308}\\
Y_{44}(\s,\t) &= \left(\s^4+16 \s^3 \t+36 \s^2 \t^2+16 \s \t^3+\t^4\right)-\frac{8}{5}  \left(\s^3+9 \s^2 \t+9 \s \t^2+\t^3\right)\\&+ \frac{4}{5}  \left(\s^2+4 \s \t+\t^2\right)+-\frac{2}{15}  (\s+\t)+\frac{1}{210}}
which can be derived from  \c{osborn}. 

The projector with $(cd)=(00)$ is relevant for the conformal block decomposition of $\la \O_p\O_p\O_p\O_p\ra$ in mean field theory, which is a sum over channels of identity exchange:
\es{disc}{\F_p^{MFT}(u,v;\s,\tau) &= 1+(\s u)^p + \left(\t {u\o v}\right)^p\\
&= 1+\sum_{a,b}Y_{ab}(\s,\tau) \Pc_{ab|00}(p)\left[u^p+(-1)^{a+b}\left({u\o v}\right)^p\right]}
The first several low-lying representations $(ab)$ in the symmetric product $[\Rc_p\otimes \Rc_p]_{\rm \, sym}$ are
\es{proj00}{\Pc_{00|00}(p) &= \frac{360}{(p+1) (p+2) (p+3)^2 (p+4) (p+5)}\\
\Pc_{11|00}(p) &= \frac{16800 p}{(p+1) (p+3)^2 (p+4)^2 (p+5) (p+6)}\\
\Pc_{20|00}(p) &=\frac{151200 (p-1) p}{(p+1) (p+2) (p+3)^2 (p+4) (p+5) (p+6) (p+7)}\\
\Pc_{22|00}(p) &= \frac{264600 (p-1) p}{(p+3)^2 (p+4)^2 (p+5)^2 (p+6) (p+7)}\\
\Pc_{31|00}(p) &=\frac{4191264 (p-2) (p-1) p}{(p+1) (p+3)^2 (p+4)^2 (p+5) (p+6) (p+7) (p+8)}\\
\Pc_{33|00}(p) &= \frac{2794176 (p-2) (p-1) p^2}{(p+3)^2 (p+4)^2 (p+5)^2 (p+6)^2 (p+7) (p+8)}\\
\Pc_{40|00}(p) &= \frac{15135120 (p-3) (p-2) (p-1) p}{(p+1) (p+2) (p+3)^2 (p+4) (p+5) (p+6) (p+7) (p+8) (p+9)}\\
\Pc_{42|00}(p) &=\frac{51891840 (p-3) (p-2) (p-1) p}{(p+3)^2 (p+4)^2 (p+5)^2 (p+6) (p+7) (p+8) (p+9)}\\
\Pc_{44|00}(p) &= \frac{23783760 (p-3) (p-2) (p-1)^2 p^2}{(p+3)^2 (p+4)^2 (p+5)^2 (p+6)^2 (p+7)^2 (p+8) (p+9)}}
The projector with $(cd)=(11)$ was needed for the conformal block decomposition of the change in $\la \O_p\O_p\O_p\O_p\ra$ after double-trace flow triggered by $\int \!\O_2^2$, where $\O_2$ sits in the $(11)$ representation. The first several low-lying representations $(ab)$ in the symmetric product $[\Rc_p\otimes \Rc_p]_{\rm \,sym}$ are
\es{proj11}{%f_{10}(p) &= \quad (\ell=1,3,\ldots)\\
\Pc_{00|11}(p) &= \frac{270 (p+6)}{p (p+1) (p+2)^2 (p+3)^2 (p+5)}\\
\Pc_{11|11}(p) &= \frac{4200 \left(3 p^4+36 p^3+100 p^2-48 p-64\right)}{p (p+1) (p+2)^2 (p+3)^2 (p+4)^2 (p+5) (p+6)}\\
\Pc_{20|11}(p) &= \frac{37800 (p-1) \left(3 p^2+18 p-56\right)}{p (p+1) (p+2)^2 (p+3)^2 (p+5) (p+6) (p+7)}\\
\Pc_{22|11}(p) &= \frac{66150 (p-1) \left(3 p^4+36 p^3+52 p^2-336 p+320\right)}{p (p+2)^2 (p+3)^2 (p+4)^2 (p+5)^2 (p+6) (p+7)}\\
\Pc_{31|11}(p) &= \frac{1047816 (p-2) (p-1) \left(3 p^4+36 p^3+28 p^2-480 p-352\right)}{p (p+1) (p+2)^2 (p+3)^2 (p+4)^2 (p+5) (p+6) (p+7) (p+8)}\\
\Pc_{33|11}(p) &= \frac{2095632 (p-2) (p-1) \left(p^4+12 p^3-4 p^2-240 p+576\right)}{(p+2)^2 (p+3)^2 (p+4)^2 (p+5)^2 (p+6)^2 (p+7) (p+8)}\\
\Pc_{40|11}(p) &= \frac{11351340 (p-3) (p-2) (p-1) \left(p^2+6 p-48\right)}{p (p+1) (p+2)^2 (p+3)^2 (p+5) (p+6) (p+7) (p+8) (p+9)}\\
\Pc_{42|11}(p) &= \frac{38918880 (p-3) (p-2) (p-1) \left(p^4+12 p^3-12 p^2-288 p+224\right)}{p (p+2)^2 (p+3)^2 (p+4)^2 (p+5)^2 (p+6) (p+7) (p+8) (p+9)}\\
\Pc_{44|11}(p) &= \frac{5945940 (p-3) (p-2) (p-1)^2 \left(3 p^4+36 p^3-92 p^2-1200 p+4928\right)}{(p+2)^2 (p+3)^2 (p+4)^2 (p+5)^2 (p+6)^2 (p+7)^2 (p+8) (p+9)}}
For both $(cd)=(00)$ and $(11)$, one check on these functions are the zeroes at $p=1,2,3$: these reflect, correctly, the absence of the $(ab)$ representations in the product $\Rc_p\otimes \Rc_p$ for $p<a$. Note the universal behavior of these projectors at large $p$, where $\sim 1/p^6$; in particular, their ratio goes to a constant. 

The above data are sufficient, using \eqr{ad1}-\eqr{ad2}, to compute $\d\g_{\ell}^{(ab)}(p)$, the change under RG flow of anomalous dimensions of $[\O_p\O_p]_{0,\ell}^{(ab)}$, for all representations appearing in the symmetric product $[\Rc_p\otimes \Rc_p]_{\rm sym}$ with $a\leq 4$. For $p=4$, this is the full set of symmetric representations; moreover, for $p=4$ and $a=3,4$, these operators are protected in the UV, so $\d\g_{\ell}^{(ab)}(4)$ equals the IR anomalous dimensions, as explained in Section \ref{s432}.

\ssec{Computing $\d\g_\ell^{(ab)}(p)$}

The starting point for this calculation is \eqr{chan}. The strategy is to decompose it into conformal blocks, picking off the terms that contain the anomalous dimensions and applying the results of \c{kirilin}. 

The second line of \eqr{chan} contains the double-trace exchanges, which have twist $\t=p+2n$ at infinite $N$. This can be inferred from the leading power of $u$: at $u\ll1$, a conformal family whose primary has twist $\t$ contributes terms of order $u^{\t/2}$ times positive integer powers. Each power of the anomalous dimension comes with a power of $\log u$. Putting these facts together, we must solve the equations
\es{appbeq}{-{16p^2\o \pi^{5/2}C_T}&\sum_{a,b} Y_{ab}(\s,\tau)\Pc_{ab|11}(p)\left[u^{p\o 2} \bar{D}_{1, {1\o 2}, 1, {1\o 2}}(u,v)+(-1)^{a+b}\left(\frac{u}{v}\right)^{p\o 2} v \bar{D}_{{1\o 2}, 1, 1, {1\o 2}}(u,v)\right]_{\log u}\\
= &\sum_{a,b}Y_{ab}(\s,\tau)\sum_{n=0}^\i u^{{p\o 2}+n} \sum_{\ell=0}^\i{1\o 2}\widetilde a^{(ab)}_{n,\ell}(p)\d\g^{(ab)}_{n,\ell}(p)\,g_{p+2n,\ell}(u,v)}
%pp2
We have employed the notation $\widetilde a^{(ab)}_{n,\ell}(p)$ for the squared OPE coefficients of MFT, and written the conformal block as
\e{}{G_{\t,\ell}(u,v) = u^{\t/2}g_{\t,\ell}(u,v)}
Let us first compute the MFT OPE coefficients $\widetilde a^{(ab)}_{n,\ell}(p)$. It follows from \eqr{disc} that they are simply those of ordinary MFT in $d=3$, in the absence of any global symmetry -- call them $a_{n,\ell}^{(0)}(p)$ -- times the $\Pc_{ab|00}(p)$ factors:
\e{b9}{\widetilde a_{n,\ell}^{(ab)} = (1+(-1)^{\ell+a+b})\Pc_{ab|00}(p) \,a_{n,\ell}^{(0)}(p)}
where \c{fkap}
\e{b10}{a_{n,\ell}^{(0)}(p) = {\left({p-1\o2}\right)_n^2\left({p\o2}\right)_{\ell+n}^2\o \ell!n!\left(\ell+{3\o2}\right)_n(p-2+n)_n(p+2n+\ell-1)_\ell\left(p+n+\ell-{3\o2}\right)_n}}
For $(ab)$ in the symmetric/anti-symmetric product of $\Rc_p\otimes \Rc_p$, only even/odd $\ell$ double-trace operators are exchanged.

Having computed the MFT result, we return to \eqr{appbeq}. We focus on the leading twist operators, with $n=0$, henceforth, and use the shorthand $\g_{\ell} \equiv \g_{0,\ell}$ and $a_{0,\ell} = a_{\ell}$. This allows us to utilize the lightcone limit,
\e{}{u\ll1~, ~~v\text{ fixed}~.}% \quad (\text{lightcone limit})}
In this limit,
\e{}{g_{p,\ell}(u\ll1, v) \approx u^{p\o 2}g^{coll}_{\ell}(v)}
where 
\e{}{g^{coll}_{\ell}(v) \equiv {}_2F_1\left({p\o 2}+\ell, {p\o 2}+\ell, p+2\ell;1-v\right)}
is the colinear, or lightcone, block. Therefore, for every $(ab)$, we must solve
\es{above}{-{16p^2\o \pi^{5/2}C_T}&{\Pc_{ab|11}(p)\o \Pc_{ab|00}(p)}\left[\bar{D}_{1, {1\o 2}, 1, {1\o 2}}(u,v)+(-1)^{a+b}\left(\frac{1}{v}\right)^{p\o 2} v \bar{D}_{{1\o 2}, 1, 1, {1\o 2}}(u,v)\right]_{\log u}\\&=\sum_{\ell=0}^\i{1\o 2}(1+(-1)^{\ell+a+b})a_{\ell}^{(0)}\d\g_{\ell}^{(ab)}(p) g^{coll}_{\ell}(v)}
%pp2
This is essentially identical to the same problem in the case where $\O_p$ is uncharged under global symmetry -- in particular, it is clear that $\d\g_\ell^{(ab)}(p)$ is simply the uncharged result, multiplied by the ratio of projectors. The uncharged problem was solved in \c{kirilin}, using the fact that $g^{coll}_{\ell}(v)$ obeys a simple orthogonality condition. The result is
%pp2
\e{}{\d\g_\ell^{(ab)}(p)  =\left( \frac{64p^2}{\pi^2C_T}{\Pc_{ab|11}(p)\o \Pc_{ab|00}(p)}\right){}_4F_3\Farg{-\ell,1,{1\o 2},p+\ell-1}{{3\o2},{p\o 2},{p\o 2}}{1}}%~, ~~\text{where}~~\ell\in2\Z_{\geq 0}}
%{1+(-1)^{\ell+a+b}\o 2}
where $\ell$ is even/odd if $a+b$ is even/odd. This result is valid for any $p\in\Z_{>0}$. Moreover, it turns out that for $p\in 2\Z$ and $\ell\in\Z_{\geq 0}$, this function simplifies tremendously:
\e{}{%{}_4F_3\left(-\ell,1,{1\o 2},p+\ell-1;{3\o2};p,p;1\right)
{}_4F_3\Farg{-\ell,1,{1\o 2},p+\ell-1}{{3\o2},{p\o 2},{p\o 2}}{1} = \sum_{k=0}^{{p-4\o 2}}{c_k\o \ell+{p\o 2}+k}\quad (p= 4,6,8,\ldots)}
where
\e{ckformapp}{c_k(p) = \frac{(p-2) \left(2-\frac{p}{2}\right)_{k} \left(\frac{p-1}{2}\right)_{k}}{(p-3) \left(\frac{5}{2}-\frac{p}{2}\right)_{k} \left(\frac{p}{2}\right)_{k}}}
This is the result \eqr{ad1}--\eqr{ckform} quoted in the main text. As explained around \eqr{mixing}, for general $(ab)\supset \Rc_p\otimes\Rc_p$ this result represents a weighted average of the change in anomalous dimensions.

\sec{Contrast with SUSY-breaking $\N=4$ SYM orbifolds}\label{appc}
For context, we give a brief historical overview of one of the most well-studied non-SUSY constructions, namely, the AdS$_5\times S^5/\Gamma$ orbifolds, dual to orbifolds of 4d $\N=4$ SYM \c{Kachru:1998ys, Lawrence:1998ja}. Non-freely-acting orbifolds $\Gamma=\Z_k$ have AdS tachyons; explicit CFT calculations at weak coupling revealed the existence of an unstable effective potential \c{tseytlin, adams}, and, later, nonzero beta functions for double-trace operators comprised of twisted sector single-trace operators \c{dym3,dym2}. The AdS tachyons were conjectured to be the strongly coupled avatars of these weak coupling phenomena \c{adams}. The freely-acting orbifolds have no tachyons in AdS, but do suffer from a non-perturbative instability \c{hop}; though an initial weak coupling calculation \c{adams} revealed no apparent instability of the effective potential, these theories were nevertheless shown to break conformality \c{dym3,dym2}. The field theory picture was tied together by \c{rastelli}, who showed that along any fixed line of $d=4$ CFTs with adjoint fields, and at any value of the marginal coupling, the CFT develops nonzero beta functions if and only if the perturbative vacuum at the origin of moduli space is unstable. No such examples survived further scrutiny. The orbifold studies were extended in \c{murugan} to AdS$_4\times S^7/\Gamma$, and to the ``skew-whiffed'' AdS$_4\times S^7$ background \c{dnp}, where global singlet marginal operators were found in both cases; this is likely to imply a breaking of conformality, but deserves further study.

Let us also highlight the ``orientifold'' CFT construction of \c{naqvi}, which does not develop nonzero double-trace beta functions \c{daughter}, and is not known to suffer from other instabilities. It should be noted, though, that this theory ceases to be conformal away from strictly infinite $N$, so it does not belong to a sequence of CFTs with a large $N$, large gap limit.

\bibliographystyle{ssg}
\bibliography{abjmbib}

\begingroup\raggedright\begin{thebibliography}{100}

\bibitem{mald}
J.~M. Maldacena, ``{The Large N limit of superconformal field theories and
  supergravity},'' {\em Adv.Theor.Math.Phys.} {\bf 2} (1998) 231--252,
  \href{http://xxx.lanl.gov/abs/hep-th/9711200}{{\tt hep-th/9711200}}.

\bibitem{gubs}
S.~Gubser, I.~R. Klebanov, and A.~M. Polyakov, ``{Gauge theory correlators from
  noncritical string theory},'' {\em Phys.Lett.} {\bf B428} (1998) 105--114,
  \href{http://xxx.lanl.gov/abs/hep-th/9802109}{{\tt hep-th/9802109}}.

\bibitem{witt}
E.~Witten, ``{Anti-de Sitter space and holography},'' {\em
  Adv.Theor.Math.Phys.} {\bf 2} (1998) 253--291,
  \href{http://xxx.lanl.gov/abs/hep-th/9802150}{{\tt hep-th/9802150}}.

\bibitem{Heemskerk:2009pn}
I.~Heemskerk, J.~Penedones, J.~Polchinski, and J.~Sully, ``{Holography from
  Conformal Field Theory},'' {\em JHEP} {\bf 10} (2009) 079,
  \href{http://xxx.lanl.gov/abs/0907.0151}{{\tt 0907.0151}}.

\bibitem{OV}
H.~Ooguri and C.~Vafa, ``{Non-supersymmetric AdS and the Swampland},''
  \href{http://xxx.lanl.gov/abs/1610.01533}{{\tt 1610.01533}}.

\bibitem{FK}
B.~Freivogel and M.~Kleban, ``{Vacua Morghulis},''
  \href{http://xxx.lanl.gov/abs/1610.04564}{{\tt 1610.04564}}.

\bibitem{cemz}
X.~O. Camanho, J.~D. Edelstein, J.~Maldacena, and A.~Zhiboedov, ``{Causality
  Constraints on Corrections to the Graviton Three-Point Coupling},'' {\em
  JHEP} {\bf 02} (2016) 020, \href{http://xxx.lanl.gov/abs/1407.5597}{{\tt
  1407.5597}}.

\bibitem{hartman}
N.~Afkhami-Jeddi, T.~Hartman, S.~Kundu, and A.~Tajdini, ``{Einstein gravity
  3-point functions from conformal field theory},''
  \href{http://xxx.lanl.gov/abs/1610.09378}{{\tt 1610.09378}}.

\bibitem{chp}
M.~S. Costa, T.~Hansen, and J.~Penedones, ``{Bounds for OPE coefficients on the
  Regge trajectory},'' \href{http://xxx.lanl.gov/abs/1707.07689}{{\tt
  1707.07689}}.

\bibitem{dnp}
M.~J. Duff, B.~E.~W. Nilsson, and C.~N. Pope, ``{The Criterion for Vacuum
  Stability in {Kaluza-Klein} Supergravity},'' {\em Phys. Lett.} {\bf 139B}
  (1984) 154--158.

\bibitem{distler}
J.~Distler and F.~Zamora, ``{Nonsupersymmetric conformal field theories from
  stable anti-de Sitter spaces},'' {\em Adv. Theor. Math. Phys.} {\bf 2} (1999)
  1405--1439, \href{http://xxx.lanl.gov/abs/hep-th/9810206}{{\tt
  hep-th/9810206}}.

\bibitem{berkooz}
M.~Berkooz and S.-J. Rey, ``{Nonsupersymmetric stable vacua of M theory},''
  {\em JHEP} {\bf 01} (1999) 014,
  \href{http://xxx.lanl.gov/abs/hep-th/9807200}{{\tt hep-th/9807200}}. [Phys.
  Lett.B449,68(1999)].

\bibitem{murugan}
A.~Murugan, {\em {Renormalization group flows in gauge-gravity duality}}.
\newblock PhD thesis, Princeton U., 2016.
\newblock \href{http://xxx.lanl.gov/abs/1610.03166}{{\tt 1610.03166}}.

\bibitem{naqvi}
A.~Armoni and A.~Naqvi, ``{A Non-Supersymmetric Large-N 3D CFT And Its Gravity
  Dual},'' {\em JHEP} {\bf 09} (2008) 119,
  \href{http://xxx.lanl.gov/abs/0806.4068}{{\tt 0806.4068}}.

\bibitem{Fischbacher:2010ec}
T.~Fischbacher, K.~Pilch, and N.~P. Warner, ``{New Supersymmetric and Stable,
  Non-Supersymmetric Phases in Supergravity and Holographic Field Theory},''
  \href{http://xxx.lanl.gov/abs/1010.4910}{{\tt 1010.4910}}.

\bibitem{Gaiotto:2009mv}
D.~Gaiotto and A.~Tomasiello, ``{The gauge dual of Romans mass},'' {\em JHEP}
  {\bf 01} (2010) 015, \href{http://xxx.lanl.gov/abs/0901.0969}{{\tt
  0901.0969}}.

\bibitem{banks}
T.~Banks, ``{Note on a Paper by Ooguri and Vafa},''
  \href{http://xxx.lanl.gov/abs/1611.08953}{{\tt 1611.08953}}.

\bibitem{ABJM}
O.~Aharony, O.~Bergman, D.~L. Jafferis, and J.~Maldacena, ``{N=6 superconformal
  Chern-Simons-matter theories, M2-branes and their gravity duals},'' {\em
  JHEP} {\bf 10} (2008) 091, \href{http://xxx.lanl.gov/abs/0806.1218}{{\tt
  0806.1218}}.

\bibitem{kirilin}
S.~Giombi, V.~Kirilin, and E.~Perlmutter, ``{to appear},''.

\bibitem{caron-huot}
S.~Caron-Huot, ``{Analyticity in Spin in Conformal Theories},''
  \href{http://xxx.lanl.gov/abs/1703.00278}{{\tt 1703.00278}}.

\bibitem{kw2}
I.~R. Klebanov and E.~Witten, ``{AdS / CFT correspondence and symmetry
  breaking},'' {\em Nucl. Phys.} {\bf B556} (1999) 89--114,
  \href{http://xxx.lanl.gov/abs/hep-th/9905104}{{\tt hep-th/9905104}}.

\bibitem{wittendt}
E.~Witten, ``{Multitrace operators, boundary conditions, and AdS / CFT
  correspondence},'' \href{http://xxx.lanl.gov/abs/hep-th/0112258}{{\tt
  hep-th/0112258}}.

\bibitem{Mueck:2002gm}
W.~Mueck, ``{An Improved correspondence formula for AdS / CFT with multitrace
  operators},'' {\em Phys. Lett.} {\bf B531} (2002) 301--304,
  \href{http://xxx.lanl.gov/abs/hep-th/0201100}{{\tt hep-th/0201100}}.

\bibitem{Berkooz:2002ug}
M.~Berkooz, A.~Sever, and A.~Shomer, ``{'Double trace' deformations, boundary
  conditions and space-time singularities},'' {\em JHEP} {\bf 05} (2002) 034,
  \href{http://xxx.lanl.gov/abs/hep-th/0112264}{{\tt hep-th/0112264}}.

\bibitem{Gubser:2002zh}
S.~S. Gubser and I.~Mitra, ``{Double trace operators and one loop vacuum energy
  in AdS / CFT},'' {\em Phys. Rev.} {\bf D67} (2003) 064018,
  \href{http://xxx.lanl.gov/abs/hep-th/0210093}{{\tt hep-th/0210093}}.

\bibitem{Gubser:2002vv}
S.~S. Gubser and I.~R. Klebanov, ``{A Universal result on central charges in
  the presence of double trace deformations},'' {\em Nucl. Phys.} {\bf B656}
  (2003) 23--36, \href{http://xxx.lanl.gov/abs/hep-th/0212138}{{\tt
  hep-th/0212138}}.

\bibitem{HartmanRastelli}
T.~Hartman and L.~Rastelli, ``{Double-trace deformations, mixed boundary
  conditions and functional determinants in AdS/CFT},'' {\em JHEP} {\bf 01}
  (2008) 019, \href{http://xxx.lanl.gov/abs/hep-th/0602106}{{\tt
  hep-th/0602106}}.

\bibitem{Diaz:2007an}
D.~E. Diaz and H.~Dorn, ``{Partition functions and double-trace deformations in
  AdS/CFT},'' {\em JHEP} {\bf 05} (2007) 046,
  \href{http://xxx.lanl.gov/abs/hep-th/0702163}{{\tt hep-th/0702163}}.

\bibitem{hks}
T.~Hartman, C.~A. Keller, and B.~Stoica, ``{Universal Spectrum of 2d Conformal
  Field Theory in the Large c Limit},'' {\em JHEP} {\bf 09} (2014) 118,
  \href{http://xxx.lanl.gov/abs/1405.5137}{{\tt 1405.5137}}.

\bibitem{mms}
J.~M. Maldacena, J.~Michelson, and A.~Strominger, ``{Anti-de Sitter
  fragmentation},'' {\em JHEP} {\bf 02} (1999) 011,
  \href{http://xxx.lanl.gov/abs/hep-th/9812073}{{\tt hep-th/9812073}}.

\bibitem{switt}
N.~Seiberg and E.~Witten, ``{The D1 / D5 system and singular CFT},'' {\em JHEP}
  {\bf 04} (1999) 017, \href{http://xxx.lanl.gov/abs/hep-th/9903224}{{\tt
  hep-th/9903224}}.

\bibitem{Nakayama:2015hga}
Y.~Nakayama and Y.~Nomura, ``{Weak gravity conjecture in the AdS/CFT
  correspondence},'' {\em Phys. Rev.} {\bf D92} (2015), no.~12 126006,
  \href{http://xxx.lanl.gov/abs/1509.01647}{{\tt 1509.01647}}.

\bibitem{adams}
A.~Adams and E.~Silverstein, ``{Closed string tachyons, AdS / CFT, and large N
  QCD},'' {\em Phys. Rev.} {\bf D64} (2001) 086001,
  \href{http://xxx.lanl.gov/abs/hep-th/0103220}{{\tt hep-th/0103220}}.

\bibitem{hop}
G.~T. Horowitz, J.~Orgera, and J.~Polchinski, ``{Nonperturbative Instability of
  AdS(5) x S**5/Z(k)},'' {\em Phys. Rev.} {\bf D77} (2008) 024004,
  \href{http://xxx.lanl.gov/abs/0709.4262}{{\tt 0709.4262}}.

\bibitem{bf}
P.~Breitenlohner and D.~Z. Freedman, ``{Positive Energy in anti-De Sitter
  Backgrounds and Gauged Extended Supergravity},'' {\em Phys. Lett.} {\bf 115B}
  (1982) 197--201.

\bibitem{cordova}
C.~Cordova, T.~T. Dumitrescu, and K.~Intriligator, ``{Multiplets of
  Superconformal Symmetry in Diverse Dimensions},''
  \href{http://xxx.lanl.gov/abs/1612.00809}{{\tt 1612.00809}}.

\bibitem{KW}
I.~R. Klebanov and E.~Witten, ``{Superconformal field theory on three-branes at
  a Calabi-Yau singularity},'' {\em Nucl. Phys.} {\bf B536} (1998) 199--218,
  \href{http://xxx.lanl.gov/abs/hep-th/9807080}{{\tt hep-th/9807080}}.

\bibitem{strassler}
M.~J. Strassler, ``{Nonsupersymmetric theories with light scalar fields and
  large hierarchies},'' \href{http://xxx.lanl.gov/abs/hep-th/0309122}{{\tt
  hep-th/0309122}}.

\bibitem{gk}
S.~S. Gubser and I.~R. Klebanov, ``{A Universal result on central charges in
  the presence of double trace deformations},'' {\em Nucl. Phys.} {\bf B656}
  (2003) 23--36, \href{http://xxx.lanl.gov/abs/hep-th/0212138}{{\tt
  hep-th/0212138}}.

\bibitem{torri}
I.~R. Klebanov and G.~Torri, ``{M2-branes and AdS/CFT},'' {\em Int. J. Mod.
  Phys.} {\bf A25} (2010) 332--350,
  \href{http://xxx.lanl.gov/abs/0909.1580}{{\tt 0909.1580}}.

\bibitem{lambert}
J.~Bagger, N.~Lambert, S.~Mukhi, and C.~Papageorgakis, ``{Multiple Membranes in
  M-theory},'' {\em Phys. Rept.} {\bf 527} (2013) 1--100,
  \href{http://xxx.lanl.gov/abs/1203.3546}{{\tt 1203.3546}}.

\bibitem{pufuf}
D.~Z. Freedman and S.~S. Pufu, ``{The holography of $F$-maximization},'' {\em
  JHEP} {\bf 03} (2014) 135, \href{http://xxx.lanl.gov/abs/1302.7310}{{\tt
  1302.7310}}.

\bibitem{petkou}
H.~Osborn and A.~C. Petkou, ``{Implications of conformal invariance in field
  theories for general dimensions},'' {\em Annals Phys.} {\bf 231} (1994)
  311--362, \href{http://xxx.lanl.gov/abs/hep-th/9307010}{{\tt
  hep-th/9307010}}.

\bibitem{pufu}
S.~M. Chester, J.~Lee, S.~S. Pufu, and R.~Yacoby, ``{The $ \mathcal{N}=8 $
  superconformal bootstrap in three dimensions},'' {\em JHEP} {\bf 09} (2014)
  143, \href{http://xxx.lanl.gov/abs/1406.4814}{{\tt 1406.4814}}.

\bibitem{Castellani:1984vv}
L.~Castellani, R.~D'Auria, P.~Fre, K.~Pilch, and P.~van Nieuwenhuizen, ``{The
  Bosonic Mass Formula for Freund-rubin Solutions of $d=11$ Supergravity on
  General Coset Manifolds},'' {\em Class. Quant. Grav.} {\bf 1} (1984)
  339--348.

\bibitem{Duff:1986hr}
M.~J. Duff, B.~E.~W. Nilsson, and C.~N. Pope, ``{Kaluza-Klein Supergravity},''
  {\em Phys. Rept.} {\bf 130} (1986) 1--142.

\bibitem{liu}
J.~T. Liu and W.~Zhao, ``{One-loop supergravity on $\mathrm{AdS}_4\times
  S^7/\mathbb{Z}_k$ and comparison with ABJM theory},'' {\em JHEP} {\bf 11}
  (2016) 099, \href{http://xxx.lanl.gov/abs/1609.02558}{{\tt 1609.02558}}.

\bibitem{GY}
G.~Gur-Ari and R.~Yacoby, ``{Three Dimensional Bosonization From
  Supersymmetry},'' {\em JHEP} {\bf 11} (2015) 013,
  \href{http://xxx.lanl.gov/abs/1507.04378}{{\tt 1507.04378}}.

\bibitem{dym3}
A.~Dymarsky, I.~R. Klebanov, and R.~Roiban, ``{Perturbative search for fixed
  lines in large N gauge theories},'' {\em JHEP} {\bf 08} (2005) 011,
  \href{http://xxx.lanl.gov/abs/hep-th/0505099}{{\tt hep-th/0505099}}.

\bibitem{dym2}
A.~Dymarsky, I.~R. Klebanov, and R.~Roiban, ``{Perturbative gauge theory and
  closed string tachyons},'' {\em JHEP} {\bf 11} (2005) 038,
  \href{http://xxx.lanl.gov/abs/hep-th/0509132}{{\tt hep-th/0509132}}.

\bibitem{rastelli}
E.~Pomoni and L.~Rastelli, ``{Large N Field Theory and AdS Tachyons},'' {\em
  JHEP} {\bf 04} (2009) 020, \href{http://xxx.lanl.gov/abs/0805.2261}{{\tt
  0805.2261}}.

\bibitem{9904017}
D.~Z. Freedman, S.~S. Gubser, K.~Pilch, and N.~P. Warner, ``{Renormalization
  group flows from holography supersymmetry and a c theorem},'' {\em Adv.
  Theor. Math. Phys.} {\bf 3} (1999) 363--417,
  \href{http://xxx.lanl.gov/abs/hep-th/9904017}{{\tt hep-th/9904017}}.

\bibitem{klose}
M.~Benna, I.~Klebanov, T.~Klose, and M.~Smedback, ``{Superconformal
  Chern-Simons Theories and AdS(4)/CFT(3) Correspondence},'' {\em JHEP} {\bf
  09} (2008) 072, \href{http://xxx.lanl.gov/abs/0806.1519}{{\tt 0806.1519}}.

\bibitem{ABJ}
O.~Aharony, O.~Bergman, and D.~L. Jafferis, ``{Fractional M2-branes},'' {\em
  JHEP} {\bf 11} (2008) 043, \href{http://xxx.lanl.gov/abs/0807.4924}{{\tt
  0807.4924}}.

\bibitem{fabbri}
D.~Fabbri, P.~Fre, L.~Gualtieri, and P.~Termonia, ``{M theory on AdS(4) x
  M**111: The Complete Osp(2|4) x SU(3) x SU(2) spectrum from harmonic
  analysis},'' {\em Nucl. Phys.} {\bf B560} (1999) 617--682,
  \href{http://xxx.lanl.gov/abs/hep-th/9903036}{{\tt hep-th/9903036}}.

\bibitem{fabbri2}
D.~Fabbri, P.~Fre', L.~Gualtieri, C.~Reina, A.~Tomasiello, A.~Zaffaroni, and
  A.~Zampa, ``{3-D superconformal theories from Sasakian seven manifolds: New
  nontrivial evidences for AdS(4) / CFT(3)},'' {\em Nucl. Phys.} {\bf B577}
  (2000) 547--608, \href{http://xxx.lanl.gov/abs/hep-th/9907219}{{\tt
  hep-th/9907219}}.

\bibitem{Martelli:2008si}
D.~Martelli and J.~Sparks, ``{Moduli spaces of Chern-Simons quiver gauge
  theories and AdS(4)/CFT(3)},'' {\em Phys. Rev.} {\bf D78} (2008) 126005,
  \href{http://xxx.lanl.gov/abs/0808.0912}{{\tt 0808.0912}}.

\bibitem{Dong1}
X.~Dong, D.~Z. Freedman, and Y.~Zhao, ``{Explicitly Broken Supersymmetry with
  Exactly Massless Moduli},'' {\em JHEP} {\bf 06} (2016) 090,
  \href{http://xxx.lanl.gov/abs/1410.2257}{{\tt 1410.2257}}.

\bibitem{Dong2}
X.~Dong, D.~Z. Freedman, and Y.~Zhao, ``{AdS/CFT and the Little Hierarchy
  Problem},'' \href{http://xxx.lanl.gov/abs/1510.01741}{{\tt 1510.01741}}.

\bibitem{witten2003}
E.~Witten, ``{SL(2,Z) action on three-dimensional conformal field theories with
  Abelian symmetry},'' \href{http://xxx.lanl.gov/abs/hep-th/0307041}{{\tt
  hep-th/0307041}}.

\bibitem{tseytlin}
A.~A. Tseytlin and K.~Zarembo, ``{Effective potential in nonsupersymmetric
  SU(N) x SU(N) gauge theory and interactions of type 0 D3-branes},'' {\em
  Phys. Lett.} {\bf B457} (1999) 77--86,
  \href{http://xxx.lanl.gov/abs/hep-th/9902095}{{\tt hep-th/9902095}}.

\bibitem{ale}
A.~Adams, J.~Polchinski, and E.~Silverstein, ``{Don't panic! Closed string
  tachyons in ALE space-times},'' {\em JHEP} {\bf 10} (2001) 029,
  \href{http://xxx.lanl.gov/abs/hep-th/0108075}{{\tt hep-th/0108075}}.

\bibitem{craps}
A.~Bernamonti and B.~Craps, ``{D-Brane Potentials from Multi-Trace Deformations
  in AdS/CFT},'' {\em JHEP} {\bf 08} (2009) 112,
  \href{http://xxx.lanl.gov/abs/0907.0889}{{\tt 0907.0889}}.

\bibitem{Klebanov:2011gs}
I.~R. Klebanov, S.~S. Pufu, and B.~R. Safdi, ``{F-Theorem without
  Supersymmetry},'' {\em JHEP} {\bf 10} (2011) 038,
  \href{http://xxx.lanl.gov/abs/1105.4598}{{\tt 1105.4598}}.

\bibitem{marino}
N.~Drukker, M.~Marino, and P.~Putrov, ``{From weak to strong coupling in ABJM
  theory},'' {\em Commun. Math. Phys.} {\bf 306} (2011) 511--563,
  \href{http://xxx.lanl.gov/abs/1007.3837}{{\tt 1007.3837}}.

\bibitem{extremal}
E.~D'Hoker, D.~Z. Freedman, S.~D. Mathur, A.~Matusis, and L.~Rastelli,
  ``{Extremal correlators in the AdS / CFT correspondence},''
  \href{http://xxx.lanl.gov/abs/hep-th/9908160}{{\tt hep-th/9908160}}.

\bibitem{0006103}
E.~D'Hoker and B.~Pioline, ``{Near extremal correlators and generalized
  consistent truncation for AdS(4|7) x S**(7|4)},'' {\em JHEP} {\bf 07} (2000)
  021, \href{http://xxx.lanl.gov/abs/hep-th/0006103}{{\tt hep-th/0006103}}.

\bibitem{bastianelli}
F.~Bastianelli and R.~Zucchini, ``{Three point functions of chiral primary
  operators in d = 3, N=8 and d = 6, N=(2,0) SCFT at large N},'' {\em Phys.
  Lett.} {\bf B467} (1999) 61--66,
  \href{http://xxx.lanl.gov/abs/hep-th/9907047}{{\tt hep-th/9907047}}.

\bibitem{Freedman:1998tz}
D.~Z. Freedman, S.~D. Mathur, A.~Matusis, and L.~Rastelli, ``{Correlation
  functions in the CFT(d) / AdS(d+1) correspondence},'' {\em Nucl. Phys.} {\bf
  B546} (1999) 96--118, \href{http://xxx.lanl.gov/abs/hep-th/9804058}{{\tt
  hep-th/9804058}}.

\bibitem{pw}
D.~Z. Freedman, K.~Pilch, S.~S. Pufu, and N.~P. Warner, ``{Boundary Terms and
  Three-Point Functions: An AdS/CFT Puzzle Resolved},'' {\em JHEP} {\bf 06}
  (2017) 053, \href{http://xxx.lanl.gov/abs/1611.01888}{{\tt 1611.01888}}.

\bibitem{Petkou:1994ad}
A.~Petkou, ``{Conserved currents, consistency relations and operator product
  expansions in the conformally invariant O(N) vector model},'' {\em Annals
  Phys.} {\bf 249} (1996) 180--221,
  \href{http://xxx.lanl.gov/abs/hep-th/9410093}{{\tt hep-th/9410093}}.

\bibitem{Petkou:2003zz}
A.~C. Petkou, ``{Evaluating the AdS dual of the critical O(N) vector model},''
  {\em JHEP} {\bf 03} (2003) 049,
  \href{http://xxx.lanl.gov/abs/hep-th/0302063}{{\tt hep-th/0302063}}.

\bibitem{Sezgin:2003pt}
E.~Sezgin and P.~Sundell, ``{Holography in 4D (super) higher spin theories and
  a test via cubic scalar couplings},'' {\em JHEP} {\bf 07} (2005) 044,
  \href{http://xxx.lanl.gov/abs/hep-th/0305040}{{\tt hep-th/0305040}}.

\bibitem{yamatsu}
N.~Yamatsu, ``{Finite-Dimensional Lie Algebras and Their Representations for
  Unified Model Building},'' \href{http://xxx.lanl.gov/abs/1511.08771}{{\tt
  1511.08771}}.

\bibitem{ferrara}
S.~Ferrara and E.~Sokatchev, ``{Universal properties of superconformal OPEs for
  1/2 BPS operators in 3 $\leq$ D $\leq$ 6},'' {\em New J. Phys.} {\bf 4}
  (2002) 2, \href{http://xxx.lanl.gov/abs/hep-th/0110174}{{\tt
  hep-th/0110174}}.

\bibitem{osborn}
M.~Nirschl and H.~Osborn, ``{Superconformal Ward identities and their
  solution},'' {\em Nucl. Phys.} {\bf B711} (2005) 409--479,
  \href{http://xxx.lanl.gov/abs/hep-th/0407060}{{\tt hep-th/0407060}}.

\bibitem{dolan}
F.~A. Dolan, L.~Gallot, and E.~Sokatchev, ``{On four-point functions of 1/2-BPS
  operators in general dimensions},'' {\em JHEP} {\bf 09} (2004) 056,
  \href{http://xxx.lanl.gov/abs/hep-th/0405180}{{\tt hep-th/0405180}}.

\bibitem{drummond}
F.~Aprile, J.~M. Drummond, P.~Heslop, and H.~Paul, ``{Unmixing Supergravity},''
  \href{http://xxx.lanl.gov/abs/1706.08456}{{\tt 1706.08456}}.

\bibitem{aldaybissi}
L.~F. Alday and A.~Bissi, ``{Loop Corrections to Supergravity on $AdS_5 \times
  S^5$},'' \href{http://xxx.lanl.gov/abs/1706.02388}{{\tt 1706.02388}}.

\bibitem{poland}
D.~Li, D.~Meltzer, and D.~Poland, ``{Non-Abelian Binding Energies from the
  Lightcone Bootstrap},'' {\em JHEP} {\bf 02} (2016) 149,
  \href{http://xxx.lanl.gov/abs/1510.07044}{{\tt 1510.07044}}.

\bibitem{fitz}
A.~L. Fitzpatrick, J.~Kaplan, D.~Poland, and D.~Simmons-Duffin, ``{The Analytic
  Bootstrap and AdS Superhorizon Locality},'' {\em JHEP} {\bf 12} (2013) 004,
  \href{http://xxx.lanl.gov/abs/1212.3616}{{\tt 1212.3616}}.

\bibitem{kz}
Z.~Komargodski and A.~Zhiboedov, ``{Convexity and Liberation at Large Spin},''
  {\em JHEP} {\bf 11} (2013) 140, \href{http://xxx.lanl.gov/abs/1212.4103}{{\tt
  1212.4103}}.

\bibitem{az}
L.~F. Alday and A.~Zhiboedov, ``{An Algebraic Approach to the Analytic
  Bootstrap},'' {\em JHEP} {\bf 04} (2017) 157,
  \href{http://xxx.lanl.gov/abs/1510.08091}{{\tt 1510.08091}}.

\bibitem{poland2}
D.~Li, D.~Meltzer, and D.~Poland, ``{Conformal Collider Physics from the
  Lightcone Bootstrap},'' {\em JHEP} {\bf 02} (2016) 143,
  \href{http://xxx.lanl.gov/abs/1511.08025}{{\tt 1511.08025}}.

\bibitem{Cornalba:2006xm}
L.~Cornalba, M.~S. Costa, J.~Penedones, and R.~Schiappa, ``{Eikonal
  Approximation in AdS/CFT: Conformal Partial Waves and Finite N Four-Point
  Functions},'' {\em Nucl. Phys.} {\bf B767} (2007) 327--351,
  \href{http://xxx.lanl.gov/abs/hep-th/0611123}{{\tt hep-th/0611123}}.

\bibitem{Fitzpatrick:2010zm}
A.~L. Fitzpatrick, E.~Katz, D.~Poland, and D.~Simmons-Duffin, ``{Effective
  Conformal Theory and the Flat-Space Limit of AdS},'' {\em JHEP} {\bf 07}
  (2011) 023, \href{http://xxx.lanl.gov/abs/1007.2412}{{\tt 1007.2412}}.

\bibitem{Fitzpatrick:2011hh}
A.~L. Fitzpatrick and D.~Shih, ``{Anomalous Dimensions of Non-Chiral Operators
  from AdS/CFT},'' {\em JHEP} {\bf 10} (2011) 113,
  \href{http://xxx.lanl.gov/abs/1104.5013}{{\tt 1104.5013}}.

\bibitem{Giombi:2017hpr}
S.~Giombi, C.~Sleight, and M.~Taronna, ``{Spinning AdS Loop Diagrams: Two Point
  Functions},'' \href{http://xxx.lanl.gov/abs/1708.08404}{{\tt 1708.08404}}.

\bibitem{GiombiYin}
S.~Giombi and X.~Yin, ``{On Higher Spin Gauge Theory and the Critical O(N)
  Model},'' {\em Phys. Rev.} {\bf D85} (2012) 086005,
  \href{http://xxx.lanl.gov/abs/1105.4011}{{\tt 1105.4011}}.

\bibitem{Giombi:2011kc}
S.~Giombi, S.~Minwalla, S.~Prakash, S.~P. Trivedi, S.~R. Wadia, and X.~Yin,
  ``{Chern-Simons Theory with Vector Fermion Matter},'' {\em Eur. Phys. J.}
  {\bf C72} (2012) 2112, \href{http://xxx.lanl.gov/abs/1110.4386}{{\tt
  1110.4386}}.

\bibitem{Aharony:2011jz}
O.~Aharony, G.~Gur-Ari, and R.~Yacoby, ``{d=3 Bosonic Vector Models Coupled to
  Chern-Simons Gauge Theories},'' {\em JHEP} {\bf 03} (2012) 037,
  \href{http://xxx.lanl.gov/abs/1110.4382}{{\tt 1110.4382}}.

\bibitem{Karch:2016sxi}
A.~Karch and D.~Tong, ``{Particle-Vortex Duality from 3d Bosonization},'' {\em
  Phys. Rev.} {\bf X6} (2016), no.~3 031043,
  \href{http://xxx.lanl.gov/abs/1606.01893}{{\tt 1606.01893}}.

\bibitem{Murugan:2016zal}
J.~Murugan and H.~Nastase, ``{Particle-vortex duality in topological insulators
  and superconductors},'' {\em JHEP} {\bf 05} (2017) 159,
  \href{http://xxx.lanl.gov/abs/1606.01912}{{\tt 1606.01912}}.

\bibitem{Seiberg:2016gmd}
N.~Seiberg, T.~Senthil, C.~Wang, and E.~Witten, ``{A Duality Web in 2+1
  Dimensions and Condensed Matter Physics},'' {\em Annals Phys.} {\bf 374}
  (2016) 395--433, \href{http://xxx.lanl.gov/abs/1606.01989}{{\tt 1606.01989}}.

\bibitem{msw}
J.~M. Maldacena, A.~Strominger, and E.~Witten, ``{Black hole entropy in M
  theory},'' {\em JHEP} {\bf 12} (1997) 002,
  \href{http://xxx.lanl.gov/abs/hep-th/9711053}{{\tt hep-th/9711053}}.

\bibitem{Larsen:1998xm}
F.~Larsen, ``{The Perturbation spectrum of black holes in N=8 supergravity},''
  {\em Nucl. Phys.} {\bf B536} (1998) 258--278,
  \href{http://xxx.lanl.gov/abs/hep-th/9805208}{{\tt hep-th/9805208}}.

\bibitem{Fujii:1998tc}
A.~Fujii, R.~Kemmoku, and S.~Mizoguchi, ``{D = 5 simple supergravity on AdS(3)
  x S**2 and N=4 superconformal field theory},'' {\em Nucl. Phys.} {\bf B574}
  (2000) 691--718, \href{http://xxx.lanl.gov/abs/hep-th/9811147}{{\tt
  hep-th/9811147}}.

\bibitem{deBoer:2009un}
J.~de~Boer, S.~El-Showk, I.~Messamah, and D.~Van~den Bleeken, ``{A Bound on the
  entropy of supergravity?},'' {\em JHEP} {\bf 02} (2010) 062,
  \href{http://xxx.lanl.gov/abs/0906.0011}{{\tt 0906.0011}}.

\bibitem{Diab:2016spb}
K.~Diab, L.~Fei, S.~Giombi, I.~R. Klebanov, and G.~Tarnopolsky, ``{On ${C}_{J}$
  and ${C}_{T}$ in the Gross Neveu and O(N) models},'' {\em J. Phys.} {\bf A49}
  (2016), no.~40 405402, \href{http://xxx.lanl.gov/abs/1601.07198}{{\tt
  1601.07198}}.

\bibitem{fkap}
A.~L. Fitzpatrick and J.~Kaplan, ``{Unitarity and the Holographic S-Matrix},''
  {\em JHEP} {\bf 10} (2012) 032, \href{http://xxx.lanl.gov/abs/1112.4845}{{\tt
  1112.4845}}.

\bibitem{Kachru:1998ys}
S.~Kachru and E.~Silverstein, ``{4-D conformal theories and strings on
  orbifolds},'' {\em Phys. Rev. Lett.} {\bf 80} (1998) 4855--4858,
  \href{http://xxx.lanl.gov/abs/hep-th/9802183}{{\tt hep-th/9802183}}.

\bibitem{Lawrence:1998ja}
A.~E. Lawrence, N.~Nekrasov, and C.~Vafa, ``{On conformal field theories in
  four-dimensions},'' {\em Nucl. Phys.} {\bf B533} (1998) 199--209,
  \href{http://xxx.lanl.gov/abs/hep-th/9803015}{{\tt hep-th/9803015}}.

\bibitem{daughter}
P.~Liendo, ``{Orientifold daughter of N=4 SYM and double-trace running},'' {\em
  Phys. Rev.} {\bf D86} (2012) 105032,
  \href{http://xxx.lanl.gov/abs/1107.3125}{{\tt 1107.3125}}.

\end{thebibliography}\endgroup

\end{document}